\documentclass[aps,prd,superscriptaddress,twoside,twocolumn,nofootinbib,10pt,%
showpacs,floatfix]{revtex4-1}

\usepackage{amsmath,amssymb}
\usepackage{graphicx,bm}
\usepackage{slashed}
\usepackage{epstopdf}
\usepackage{ulem} 
\usepackage[usenames]{color}
\usepackage{float}
\usepackage{multirow}

\renewcommand\sout{\bgroup \color{red} \ULdepth=-.5ex \ULset}

\begin{document}
\title{Investigation on the stabilities of doubly heavy tetraquark states}

\author{Woosung Park}\affiliation{Department of Physics and Institute of Physics and Applied Physics, Yonsei University, Seoul 03722, Korea}
\author{Sungsik Noh}\affiliation{Division of Science Education, Kangwon National University, Chuncheon 24341, Korea}
\date{\today}
\begin{abstract}
In our recent work\cite{Noh:2023zoq}, the mass and binding energy of $T_{cc}$ are found to be $3873$ MeV and $-2$ MeV, respectively, which align with the observations reported at LHCb\cite{LHCb:2021vvq}.
Based on our latest quark model approach, we extend our search for other potentially stable configurations of doubly heavy tetraquarks using our nonrelativistic quark model described in Ref.~\cite{Noh:2023zoq}.
Our numerical calculations indicate that the $\bar{u}\bar{s}cb$ configuration is deeply bound. However, the $\bar{u}\bar{d}cb$ configuration with the isospin symmetry in the light quark sector is relatively less bound.
In this study, we emphasize a compulsory aspect for requiring a complete set of three dimensional harmonic oscillator bases through the discussion of the $\bar{u}\bar{s}cb$ configuration and investigate the essential differences between the $\bar{u}\bar{s}cb$ and $\bar{u}\bar{d}cb$ configurations.
\end{abstract}


\maketitle

\section{Introduction}

Since the discovery of $X(3872)$\cite{Belle:2003nnu,CDF:2003cab}, the recently observed $T_{cc}$ state\cite{LHCb:2021vvq} has further increased the possibility of existence of other exotic particles.
Indeed, the identification of exotic particles such as tetraquarks\cite{CDF:2009jgo,BESIII:2013ris,Belle:2013shl} and pentaquarks\cite{LHCb:2015yax,LHCb:2019kea} gives rise to a considerable interest in the exotic hadrons.

Although experimental studies on exotic particles containing $b$ quarks are highly limited, theoretical investigations of these exotic particles are more important than ever before.
For instance, the ongoing debate over the structure of the $X(3872)$ and $T_{cc}$ highlights that understanding tetraquark structures heavier than the $T_{cc}$ could provide the theoretical foundation necessary to resolve such disputes.
The debate regarding the internal structures of $X(3872)$ and $T_{cc}$ particularly reflects these points, as the heavy quarks of the $X(3872)$ are composed of hidden charm quarks, whereas those of the $T_{cc}$ are open charm quarks.
If $X(3872)$ is a loosely bound molecular state, understanding the nature of the interactions that permit the existence of this structure is extremely crucial.
In addition, developing a theoretical framework to explain the binding mechanism of $T_{cc}$ is a significant challenge.
Therefore, theoretical investigations of other tetraquarks are necessary.
Moreover, theoretical approaches that can align with experimental results are the cornerstone of these efforts.

On the other hand, prior to the discovery of $T_{cc}$, various theoretical approaches within nonrelativistic quark models based on QCD were applied to understand the structure of $T_{cc}$\cite{Silvestre-Brac:1993zem,Semay:1994ht,Vijande:2007rf,Karliner:2017qjm,Eichten:2017ffp,Park:2018wjk,Hernandez:2019eox,Meng:2020knc,Noh:2021lqs}. Among these, it is remarkable that the bound state of $T_{cc}$ is likely to be found when its calculation is performed using a highly precise approximation method.

Conspicuously, only the works by J. Vijande\cite{Vijande:2007rf,Hernandez:2019eox} and Q. Meng and E. Hiyama\cite{Meng:2020knc,Meng:2021agn} reported a bound state for $T_{cc}$.
In Ref.~\cite{Vijande:2007rf}, the authors obtained a binding energy of $-6$ MeV for $T_{cc}$ by applying the hyperspherical harmonics formalism.
In Ref.~\cite{Hernandez:2019eox}, a binding energy of $-13$ MeV was yielded by using a generalized Gaussian variational(GGV) approach with a wave function expanded in terms of $N$ generalized Gaussians.
In Ref~\cite{Meng:2020knc,Meng:2021agn}, the binding energy was evaluated at $-23$ MeV through the Gaussian Expansion Method\cite{Hiyama:2003cu,Hiyama:2012sma}, developed by E. Hiyama, Y. Kino, and M. Kamimura.

In contrast to the above works related to the $T_{cc}$, our nonrelativistic quark model\cite{Noh:2021lqs,Noh:2023zoq} employs a complete set of three dimensional harmonic oscillator bases, which is essential for ensuring the convergence of solution to the Schr$\ddot{\rm o}$dinger equation describing a multiquark state.
One of the most important results from our recent study\cite{Noh:2023zoq} is that adoption of a Yukawa type for the hyperfine potential leads to interesting results: the $T_{cc}$ state is obtained as a weakly bound state with a mass of 3872 MeV and a binding energy of $-2$ MeV, closely aligning with experimental observations reported in Ref.~\cite{LHCb:2021vvq}.

To understand the structures of tetraquark states that meet our motivation, we conduct a comprehensive search for possible bound states among other candidates for doubly heavy tetraquarks.
By analyzing different quark configurations and their interactions quantitatively, we aim to provide deeper insights into the characteristics and stability of these exotic states.

This paper is organized as follows: Sec.~\ref{ModelDesc} provides our nonrelativistic quark model. In Sec.~\ref{ResultsAnalysis}, we present the numerical results for tetraquarks and analysis.
In our calculations with a high precision, we also emphasize a compulsory aspect for requiring a complete set of three dimensional harmonic oscillator bases by comparing these results with those obtained using the most dominant single spatial basis.
Additionally, we address a significant distinction arising from flavor symmetry breaking in the light quark sector of doubly heavy tetraquark states. Specifically, we compare the $\bar{u}\bar{d}cb$ with $(I,S)=(0,0)$ against the $\bar{u}\bar{s}cb$ with $(I,S)=(0,0)$.

\section{Model description}
\label{ModelDesc}
In our nonrelativisitc quark model, tetraquark systems are described by the following Hamiltonian\cite{Noh:2023zoq}:
\begin{eqnarray}
H &=& \sum^{4}_{i=1} \left( m_i+\frac{{\mathbf p}^{2}_i}{2 m_i} \right)-\frac{3}{4}\sum^{4}_{i<j}\frac{\lambda^{c}_{i}}{2} \,\, \frac{\lambda^{c}_{j}}{2} \left( V^{C}_{ij} + V^{CS}_{ij} - D \right),
\nonumber \\
\label{Hamiltonian}
\end{eqnarray}
where the confinement and the Yukawa type hyperfine potentials are given as follow:
\begin{eqnarray}
V^{C}_{ij} &=& - \frac{\kappa}{r_{ij}} + \frac{r_{ij}}{a^2_0},
\label{ConfineP}
\end{eqnarray}
\begin{eqnarray}
V^{CS}_{ij} &=& \frac{\hbar^2 c^2 \kappa'}{m_i m_j c^4} \frac{e^{- r_{ij} / r_{0ij}}}{(r_{0ij}) r_{ij}} \boldsymbol{\sigma}_i \cdot \boldsymbol{\sigma}_j.
\label{CSP}
\end{eqnarray}
Here, $m_i$ represents the quark mass, $\lambda^c_{i}/2$ and $\boldsymbol{\sigma}_i$ denote the SU(3) color and SU(2) spin operators, respectively, for the $i$th quark.
$r_{ij}\equiv|{\mathbf r}_i - {\mathbf r}_j |$ denotes the relative distance between the $i$th and $j$th quarks. Additionally, $\kappa'$ and $r_{0ij}$ contain additional mass dependences as follows:
\begin{eqnarray}
r_{0ij} &=& 1/ \left( \alpha + \beta \frac{m_i m_j}{m_i + m_j} \right)	,	
\label{Parameter1}
\\
\kappa' &=& \kappa_0 \left( 1+ \gamma \frac{m_i m_j}{m_i + m_j} \right)	.	
\label{Parameter2}
\end{eqnarray}
We adopt the model fitting parameters provided in Ref.~\cite{Noh:2023zoq} as follows:
\begin{eqnarray}
&\kappa=97.7 \, \textrm{MeV fm}, \quad a_0=0.0327338 \, \textrm{(MeV$^{-1}$fm)$^{1/2}$},& \nonumber \\
&D=959  \, \textrm{MeV}, & \nonumber \\
&m_{u}=315 \, \textrm{MeV}, \qquad m_{s}=610 \, \textrm{MeV}, &\nonumber \\
&m_{c}=1895 \, \textrm{MeV}, \qquad m_{b}=5274 \, \textrm{MeV},	&\nonumber \\
&\alpha = 1.1349 \, \textrm{fm$^{-1}$}, \,\, \beta = 0.0011554 \, \textrm{(MeV fm)$^{-1}$}, &	\nonumber \\
&\gamma = 0.001370 \, \textrm{MeV$^{-1}$}, \,\, \kappa_0=213.244 \, \textrm{MeV}. 	 &
\label{FitParameters}
\end{eqnarray}
The fitting results for mesons and baryons are presented in Tables~\ref{mesons} and \ref{baryons}.

\begin{table}[ht]

\caption{This table is taken directly from Ref.~\cite{Noh:2023zoq} and shows the masses of mesons obtained(Column 3) from the model calculation using the fitting parameter set given in Eq.~(\ref{FitParameters}). Column 4 shows the variational parameter $a$.}	

\centering

\begin{tabular}{cccc}
\hline
\hline
			&	Experimental	&	Mass		&	Variational		\\
Mesons		&	Value (MeV)	&	(MeV)	&	Parameter (${\rm fm}^{-2}$)	\\
\hline 
$D$			&	1864.8		&	1865.0		&	$a$ = 6.8	\\
$D^*$		&	2007.0		&	2009.4    	&	$a$ = 5.1	\\
$\eta_{c}$	&	2983.6		&	3008.2		&	$a$ = 22.4	\\
$J/\Psi$		&	3096.9		&	3131.2		&	$a$ = 17.0	\\
$D_s$		&	1968.3		&	1967.0		&	$a$ = 11.0	\\
$D^*_s$		&	2112.1		&	2103.5		&	$a$ = 5.1 	\\
$K$			&	493.68		&	497.17		&	$a$ = 7.2 	\\
$K^*$		&	891.66	 	&	872.74		&	$a$ = 3.6   	\\

$B$			&	5279.3	 	&	5276.3		&	$a$ = 6.5 	\\
$B^*$		&	5325.2		&	5331.8		&	$a$ = 5.8 	\\
$\eta_b$		&	9398.0		&	9368.6		&	$a$ = 78.9 	\\
$\Upsilon$	&	9460.3		&	9485.0		&	$a$ = 61.6 	\\
$B_s$		&	5366.8		&	5345.6		&	$a$ = 11.5 	\\
$B_s^*$		&	5415.4		&	5404.8		&	$a$ = 10.1 	\\
$B_c$		&	6275.6		&	6276.2		&	$a$ = 32.7 	\\
$B_c^*$		&		...		&	6361.8		&	$a$ = 26.9 	\\
\hline 
\hline
\label{mesons}
\end{tabular}
\end{table}

\begin{table}[ht]
\caption{Similar to Table~\ref{mesons} but for baryons, which is also taken directly from Ref.~\cite{Noh:2023zoq}. Column 4 shows the variational parameters, $a_1$ and $a_2$.}

\centering

\begin{tabular}{cccc}
\hline
\hline
				&	Experimental	&	Mass		&	\quad Variational\quad	\\
Baryons			&	Value (MeV)	&	(MeV)	&	\quad Parameters (${\rm fm}^{-2}$)\quad	\\
\hline  
$\Lambda$		&	1115.7	&	1105.1	&	\quad$a_1$ = 4.0, $a_2$ = 3.2 \quad	\\
$\Lambda_{c}$	&	2286.5	&	2268.3 	&	\quad$a_1$ = 4.2, $a_2$ = 3.8 \quad	\\
$\Xi_{cc}$		&	3621.4	&	3623.8 	&	\quad$a_1$ = 10.2, $a_2$ = 4.1 \quad	\\
$\Lambda_b$		&	5619.4	&	5600.8 	&	\quad$a_1$ = 4.3, $a_2$ = 4.2 \quad	\\
$\Sigma_{c}$		&	2452.9	&	2448.2 	&	\quad$a_1$ = 2.6, $a_2$ = 5.1 \quad	\\
$\Sigma_{c}^*$	&	2517.5	&	2530.6 	&	\quad$a_1$ = 2.3, $a_2$ = 4.4 \quad	\\
$\Sigma_{b}$		&	5811.3	&	5817.3 	&	\quad$a_1$ = 2.5, $a_2$ = 5.4 \quad	\\
$\Sigma_{b}^*$	&	5832.1	&	5849.4 	&	\quad$a_1$ = 2.4, $a_2$ = 5.1 \quad	\\
$\Sigma$			&	1192.6	&	1193.5 	&	\quad$a_1$ = 2.7, $a_2$ = 4.5 \quad	\\
$\Sigma^*$		&	1383.7	&	1400.8 	&	\quad$a_1$ = 2.0, $a_2$ = 3.0 \quad	\\
$\Xi$			&	1314.9	&	1314.3 	&	\quad$a_1$ = 4.1, $a_2$ = 4.1 \quad	\\
$\Xi^*$			&	1531.8	&	1530.4 	&	\quad$a_1$ = 3.8, $a_2$ = 2.5 \quad	\\

$\Xi_{c}$		&	2467.8	&	2465.4	&	\quad$a_1$ = 4.6, $a_2$ = 5.4 \quad	\\
$\Xi_{c}^*$		&	2645.9	&	2642.8	&	\quad$a_1$ = 2.9, $a_2$ = 5.7 \quad	\\
$\Xi_{b}$		&	5787.8	&	5785.7	&	\quad$a_1$ = 4.7, $a_2$ = 6.3 \quad	\\
$\Xi_{b}^*$		&	5945.5	&	5950.2	&	\quad$a_1$ = 3.1, $a_2$ = 6.9 \quad	\\

$p$				&	938.27	&	941.58	&	\quad$a_1$ = 2.6, $a_2$ = 2.6 \quad	\\
$\Delta$			&	1232		&	1262.8	&	\quad$a_1$ = 1.7, $a_2$ = 1.7 \quad	\\
\hline 
\hline
\label{baryons}
\end{tabular}
\end{table}
The relative coordinates sets used to solve the Schr$\ddot{\rm o}$dinger equation in the center of mass frame are provided in Ref.~\cite{Noh:2023zoq} as follows:
\begin{itemize}
	\item{Coordinate set 1}
	\begin{eqnarray}
	& \mathbf{x}_1 = \frac{1}{\sqrt{2}}({\mathbf r}_1 - {\mathbf r}_2), \qquad \mathbf{x}_2 = \frac{1}{\sqrt{2}}({\mathbf r}_3 - {\mathbf r}_4)\,, &	\nonumber
	\\
	& \mathbf{x}_3 = \frac{1}{\mu} \left( \frac{m_1 {\mathbf r}_1 + m_2 {\mathbf r}_2}{m_1 + m_2} - \frac{m_3 {\mathbf r}_3 + m_4 {\mathbf r}_4}{m_3 + m_4} \right)\,, &
\label{CoordSet1}
	\end{eqnarray}
	\item{Coordinate set 2}
	\begin{eqnarray}
	& \mathbf{y}_1 = \frac{1}{\sqrt{2}}({\mathbf r}_1 - {\mathbf r}_3), \qquad \mathbf{y}_2 = \frac{1}{\sqrt{2}}({\mathbf r}_4 - {\mathbf r}_2)\,, &	\nonumber
	\\
	& \mathbf{y}_3 = \frac{1}{\mu} \left( \frac{m_1 {\mathbf r}_1 + m_2 {\mathbf r}_3}{m_1 + m_2} - \frac{m_3 {\mathbf r}_2 + m_4 {\mathbf r}_4}{m_3 + m_4} \right)\,, &
\label{CoordSet2}
	\end{eqnarray}
	\item{Coordinate set 3}
	\begin{eqnarray}
	& \mathbf{z}_1 = \frac{1}{\sqrt{2}}({\mathbf r}_1 - {\mathbf r}_4), \qquad \mathbf{z}_2 = \frac{1}{\sqrt{2}}({\mathbf r}_2 - {\mathbf r}_3)\,, &	\nonumber
	\\
	& \mathbf{z}_3 = \frac{1}{\mu} \left( \frac{m_1 {\mathbf r}_1 + m_2 {\mathbf r}_4}{m_1 + m_2} - \frac{m_3 {\mathbf r}_2 + m_4 {\mathbf r}_3}{m_3 + m_4} \right)\,, &
\label{CoordSet3}
	\end{eqnarray}
\end{itemize}
where
\begin{eqnarray}
\mu &=& \left[ \frac{m_1^2 + m_2^2}{(m_1 + m_2)^2} + \frac{m_3^2 + m_4^2}{(m_3 + m_4)^2} \right]^{1/2} \,,	\nonumber
\end{eqnarray}
and
\begin{eqnarray}
&m_d=m_u;&	\nonumber
\\
&m_1=m_2=m_u,	\,\,	m_3=m_4=m_b& 	\quad {\rm for} \,\, \bar{u}\bar{d}bb,	\nonumber
\\
&m_1=m_u, \,\, m_2=m_s,	\,\,	m_3=m_4=m_b& 	\quad {\rm for} \,\, \bar{u}\bar{s}bb,	\nonumber
\\
&m_1=m_2=m_u,	\,\,	m_3=m_c, \,\, m_4=m_b& 	\quad {\rm for} \,\, \bar{u}\bar{d}cb,	\nonumber
\\
&m_1=m_u, \,\, m_2=m_s,	\,\,	m_3=m_c, \,\, m_4=m_b& 	\quad {\rm for} \,\, \bar{u}\bar{s}cb,	\nonumber
\\
&m_1=m_u, \,\, m_2=m_s,	\,\,	m_3=m_4=m_c& 	\quad {\rm for} \,\, \bar{u}\bar{s}cc.	\nonumber
\end{eqnarray}
Each term of the Hamiltonian can then be calculated using the orthogonal transformations between the coordinate sets presented in Eqs.~(\ref{CoordSet1})-(\ref{CoordSet3}).

In the center of mass frame, the kinetic energy denoted as $T_c$ can be described in terms of reduced masses $m_i'$ and the Jacobi coordinates in Eq.~(\ref{CoordSet1}):
\begin{eqnarray}
T_c
&=& \sum^4_{i=1} \frac{{\mathbf p}^2_i}{2m_i} - \frac{{\mathbf p}^2_{rC}}{2M}
=\frac{{\mathbf p}^2_{\mathbf{x}_1}}{2m'_1} + \frac{{\mathbf p}^2_{\mathbf{x}_2}}{2m'_2} + \frac{{\mathbf p}^2_{\mathbf{x}_3}}{2m'_3} \,,
\end{eqnarray}
where the reduced masses $m_i'$ are defined for each tetraquark configuration as follows:
\begin{eqnarray}
&&m'_1=m_u, \,\, m'_2=m_b, \,\, m'_3=\frac{2 m_u m_b}{m_u + m_b}	\quad\qquad {\rm for} \,\, \bar{u}\bar{d}bb,
\nonumber \\
&&m'_1=\frac{2 m_u m_s}{m_u + m_s}, \,\, m'_2=m_b, 	\nonumber \\
&&\hspace{0.28cm} m'_3=\frac{(3 m_u^2 + 2 m_u m_s + 3 m_s^2) m_b}{(m_u + m_s) (m_u + m_s + 2 m_b)}  \,\,\quad\qquad {\rm for} \,\, \bar{u}\bar{s}bb,
\nonumber \\
&&m'_1=m_u, \,\, m'_2=\frac{2 m_c m_b}{m_c + m_b}, \nonumber \\
&&\hspace{0.28cm} m'_3=\frac{(3 m_c^2 + 2 m_c m_b + 3 m_b^2) m_u}{(m_c + m_b) (2 m_u + m_c + m_b)}  \,\,\quad\qquad {\rm for} \,\, \bar{u}\bar{d}cb,
\nonumber \\
&&m'_1=\frac{2 m_u m_s}{m_u + m_s}, \,\, m'_2=\frac{2 m_c m_b}{m_c + m_b}, 	\nonumber \\
&&\hspace{0.28cm} m'_3=\frac{(m_u + m_s)(m_c + m_b)}{m_u + m_s + m_c + m_b}
\nonumber \\
&&\hspace{1.5cm} \times \left[ \frac{m_s^2 + m_b^2}{\left( m_s + m_b \right)^2} + \frac{m_u^2 + m_c^2}{\left( m_u + m_c \right)^2} \right]  \,\,\,\,\, {\rm for} \,\, \bar{u}\bar{s}cb,
\nonumber
\end{eqnarray}
\begin{eqnarray}
&&m'_1=\frac{2 m_u m_s}{m_u + m_s}, \,\, m'_2=m_c, 	\nonumber \\
&&\hspace{0.28cm} m'_3=\frac{(3 m_u^2 + 2 m_u m_s + 3 m_s^2) m_c}{(m_u + m_s) (m_u + m_s + 2 m_c)}  \,\,\quad\qquad {\rm for} \,\, \bar{u}\bar{s}cc.	
\nonumber
\\
\label{ReducedM}
\end{eqnarray}

It is convenient to employ the concept of quanta, $Q$, because it is useful for identifying the convergence of the tetraquark masses.
In the extreme case where all constituent quarks and all variational parameters are identical, respectively, the diagonal component of the kinetic energy is given as:
\begin{eqnarray}
\hspace{-0.5cm}\left\langle T_c \right\rangle
=
\frac{\hbar^2 c^2 a}{m} \bigg[ 2 n_1 + l_1 + 2 n_2 + l_2 + 2 n_3 + l_3 + \frac{9}{2} \bigg].
\end{eqnarray}
Therefore, in this extreme case, the kinetic energy is identical for any combination of the internal quantum numbers $\left(n_1, n_2, n_3, l_1, l_2, l_3 \right)$ as long as the sum $\left(Q \equiv 2 n_1 + 2 n_2 + 2 n_3 + l_1 + l_2 + l_3 \right)$ is unchanged. Consequently, the expectation value of the kinetic energy increases with increasing $Q$ value. We use the value of $Q$, defined in our previous works\cite{Noh:2021lqs,Park:2023ygm}.

In calculating the Hamiltonian, the total wave function is constructed by the multiplication of the spatial wave function and the color-spin(CS) basis. On the other hand, the total wave function used in calculating the masses of tetraquarks with the variational method should satisfy the symmetry properties imposed by the Pauli principle.
The methods for constructing the total wave functions of the Hamiltonian, with detailed descriptions of the CS part of the wave function, are provided in Refs.~\cite{Noh:2023zoq,Park:2018wjk,Noh:2021lqs}.

For the spatial part of the wave function, we use a complete set of harmonic oscillator bases up to the 5th quanta ($Q=8$), which is composed of totally 210 spatial bases. This is based on our previous work\cite{Noh:2021lqs}, where the ground state converges sufficiently within a few MeV.
A few spatial bases, denoted as $\psi^{spatial}_{[n_1,n_2,n_3,l_1,l_2,l_3]}$, for the tetraquarks in the $S$-wave state are presented in Eq.~(\ref{Spatial_Wave}).
Then, the masses of tetraquarks are calculated through the variational method, which determines the parameters $a_1$, $a_2$, and $a_3$ appearing in Eq.~(\ref{Spatial_Wave}).
This approach is to minimize the eigenvalue of the diagonalized Hamiltonian represented in terms of the full wave functions by optimizing the variational parameters.
The variational parameters for the masses of mesons and baryons listed in Tables~\ref{mesons}, \ref{baryons} are determined in the same way.
\begin{widetext}
\begin{eqnarray}
\psi^{spatial}_{[0,0,0,0,0,0]} &=& \left( \frac{2}{\pi} \right)^{\frac{9}{4}} a_1^{\frac{3}{4}} a_2^{\frac{3}{4}} a_3^{\frac{3}{4}} \exp \big[- a_1 \mathbf{x}_1^2 - a_2 \mathbf{x}_2^2 - a_3 \mathbf{x}_3^2 \big],
\nonumber \\
\psi^{spatial}_{[0,0,0,1,1,0]} &=& - \frac{4}{\sqrt{3}} \left( \frac{2}{\pi} \right)^{\frac{9}{4}} a_1^{\frac{5}{4}} a_2^{\frac{5}{4}} a_3^{\frac{3}{4}} \mathbf{x}_1 \cdot \mathbf{x}_2 \exp \big[- a_1 \mathbf{x}_1^2 - a_2 \mathbf{x}_2^2 - a_3 \mathbf{x}_3^2 \big],
\nonumber \\
\psi^{spatial}_{[0,0,0,1,0,1]} &=& - \frac{4}{\sqrt{3}} \left( \frac{2}{\pi} \right)^{\frac{9}{4}} a_1^{\frac{5}{4}} a_2^{\frac{3}{4}} a_3^{\frac{5}{4}} \mathbf{x}_1 \cdot \mathbf{x}_3 \exp \big[- a_1 \mathbf{x}_1^2 - a_2 \mathbf{x}_2^2 - a_3 \mathbf{x}_3^2 \big],
\nonumber \\
\psi^{spatial}_{[0,0,0,0,1,1]} &=& - \frac{4}{\sqrt{3}} \left( \frac{2}{\pi} \right)^{\frac{9}{4}} a_1^{\frac{3}{4}} a_2^{\frac{5}{4}} a_3^{\frac{5}{4}} \mathbf{x}_2 \cdot \mathbf{x}_3 \exp \big[- a_1 \mathbf{x}_1^2 - a_2 \mathbf{x}_2^2 - a_3 \mathbf{x}_3^2 \big],
\nonumber \\
\psi^{spatial}_{[0,0,0,2,2,0]} &=& - \frac{8}{3 \sqrt{5}} \left( \frac{2}{\pi} \right)^{\frac{9}{4}} a_1^{\frac{7}{4}} a_2^{\frac{7}{4}} a_3^{\frac{3}{4}} \bigg[ \mathbf{x}_1^2 \mathbf{x}_2^2 -3 \left(\mathbf{x}_1 \cdot \mathbf{x}_2 \right)^2 \bigg] \exp \big[- a_1 \mathbf{x}_1^2 - a_2 \mathbf{x}_2^2 - a_3 \mathbf{x}_3^2 \big],
\nonumber \\
\psi^{spatial}_{[0,0,0,1,2,3]} &=& \frac{32}{15 \sqrt{21}} \left( \frac{2}{\pi} \right)^{\frac{9}{4}} a_1^{\frac{5}{4}} a_2^{\frac{7}{4}} a_3^{\frac{9}{4}}
\nonumber \\
&&
\times \bigg[ 6 \mathbf{x}_3^2 \left(\mathbf{x}_1 \cdot \mathbf{x}_2 \right) \left(\mathbf{x}_2 \cdot \mathbf{x}_3 \right) + 3 \left(\mathbf{x}_1 \cdot \mathbf{x}_3 \right) \left( \mathbf{x}_2^2 \mathbf{x}_3^2 - 5 \left(\mathbf{x}_2 \cdot \mathbf{x}_3 \right)^2 \right) \bigg] \exp \big[- a_1 \mathbf{x}_1^2 - a_2 \mathbf{x}_2^2 - a_3 \mathbf{x}_3^2 \big],
\nonumber \\
\psi^{spatial}_{[0,0,0,2,2,4]} &=& \frac{32 \sqrt{2}}{105 \sqrt{3}} \left( \frac{2}{\pi} \right)^{\frac{9}{4}} a_1^{\frac{7}{4}} a_2^{\frac{7}{4}} a_3^{\frac{11}{4}}
\nonumber \\
&&
\times \bigg[ \mathbf{x}_1^2 \mathbf{x}_2^2 \mathbf{x}_3^4 + 2 \mathbf{x}_3^4 \left(\mathbf{x}_1 \cdot \mathbf{x}_2 \right)^2 - 20 \mathbf{x}_3^2 \left(\mathbf{x}_1 \cdot \mathbf{x}_2 \right) \left(\mathbf{x}_1 \cdot \mathbf{x}_3 \right) \left(\mathbf{x}_2 \cdot \mathbf{x}_3 \right) + 35 \left(\mathbf{x}_1 \cdot \mathbf{x}_3 \right)^2 \left(\mathbf{x}_2 \cdot \mathbf{x}_3 \right)^2 
\nonumber \\
&&
\hspace{0.5cm} - 5 \left( \mathbf{x}_2^2 \mathbf{x}_3^2 \left(\mathbf{x}_1 \cdot \mathbf{x}_3 \right)^2 + \mathbf{x}_1^2 \mathbf{x}_3^2 \left(\mathbf{x}_2 \cdot \mathbf{x}_3 \right)^2\right) \bigg] \exp \big[- a_1 \mathbf{x}_1^2 - a_2 \mathbf{x}_2^2 - a_3 \mathbf{x}_3^2 \big].
\label{Spatial_Wave}
\end{eqnarray}
\end{widetext}

\section{Numerical Results and Analysis}
\label{ResultsAnalysis}

In our earlier work presented in Ref.~\cite{Noh:2023zoq}, it is found that the $\bar{u}\bar{d}bb$ with $(I,S) = (0,1)$ state is deeply bound, whereas the $\bar{u}\bar{d}cc$ with $(I,S) = (0,1)$ state is only slightly bound.
The binding energies of the $\bar{u}\bar{d}bb$ and $\bar{u}\bar{d}cc$ are $-144$ MeV and $-2$ MeV, respectively.
These two tetraquark states can be used as a starting point for establishing a certain standard against which the stability of doubly heavy tetraquarks can be measured.
In addition, the $\bar{u}\bar{d}cb$ with $(I,S) = (0,1)$ is also found to be a bound state with a binding energy of $-18$ MeV\cite{Noh:2023zoq}.

For this reason, we investigate various tetraquark configurations containing two heavy quarks and the results are presented in Table~\ref{result}.
As shown in Table~\ref{result}, while the $\bar{u}\bar{d}bb$ with $(I,S) = (0,1)$ state is strongly bound, the same quark configuration with $(I,S) = (0,0)$ is not stable.
The expectation value of the kinetic energy is lowest for the spatial wave function of the first quanta, $\psi^{spatial}_{[0,0,0,0,0,0]}$.
This spatial wave function contributes most significantly to the ground states in the $S$ wave\cite{Noh:2021lqs}, which is symmetric under permutations (12) and (34), respectively.

For the $\bar{u}\bar{d}bb$ with $I=0$ state, the full wave function should be symmetric under permutation (12) and antisymmetric under permutation (34), since the isospin part is antisymmetric under permutation (12).
The spatial wave function with the lowest quantum numbers is included in the calculation of the $\bar{u}\bar{d}bb$ with $(I,S) = (0,1)$ state because there exist two CS states that are symmetric under permutation (12) and antisymmetric under (34).
However, for the $\bar{u}\bar{d}bb$ with $(I,S) = (0,0)$ state, this spatial wave function is excluded because there is no CS states satisfying the Pauli principle in this case, unlike the $\bar{u}\bar{d}bb$ with $(I,S) = (0,1)$ state.
For the $\bar{u}\bar{d}bb$ with $(I,S) = (0,0)$ state, the next lowest spatial wave function satisfying the Pauli principle is the third one, $\psi^{spatial}_{[0,0,0,1,0,1]}$, in Eq.~(\ref{Spatial_Wave}).
Therefore, it is inevitable that the $\bar{u}\bar{d}bb$ with $(I,S) = (0,0)$ state cannot be bound against the lowest threshold mesons $BB$.
For the same reason, the $\bar{u}\bar{d}bb$ with $(I,S) = (0,2)$ state cannot be bound.

On the other hand, it is remarkable that the $\bar{u}\bar{s}cb$ with $(I,S) = (1/2,0)$ state is the most deeply bound state, second only to the $\bar{u}\bar{d}bb$ with $(I,S) = (0,1)$ state.
Given that the $\bar{u}\bar{s}cb$ state can provide critical insights into the structure of heavy tetraquarks containing $c$ and $b$ quarks, it is essential to discuss the factors contributing to its strong binding.
To begin with, our numerical analysis will focus on the $\bar{u}\bar{s}cb$ with $(I,S) = (1/2,0)$ state.

\begin{widetext}

\begin{table}[h]
\caption{Masses and binding energies $B_{T}$ of the tetraquark states. The binding energy $B_{T}$ is defined as the difference between the tetraquark mass and the sum of the masses of the lowest threshold mesons, $B_{T} \equiv M_{Tetraquark} - M_{meson 1} - M_{meson 2}$. The lowest threshold mesons are presented in column 3, without specifying the antiparticle symbols. Results for the $\bar{u}\bar{d}cb$ and $\bar{u}\bar{d}bb$ with $(I,S)=(0,1)$ states are taken from Table~III of Ref.~\cite{Noh:2023zoq}.}
\begin{tabular}{ccccccccccc}
\hline
\hline
Configuration		&&	$(I,S)$	&&	Thresholds	&&	Mass (MeV)	&&  Variational Parameters (fm$^{-2}$)    &&  $B_{T}$ (MeV)	\\
\hline 
$\bar{u}\bar{d}bb$	&&	$(0,0)$	&&	$BB$	&&	11001   &&  $a_1 = 2.7$, $a_2 = 9.2$, $a_3 = 2.9$	   &&	+442\\
$\bar{u}\bar{d}bb$	&&	$(0,1)$	&&	$BB^*$&&	10464   &&  $a_1 = 3.7$, $a_2 = 19.6$, $a_3 = 3.6$	   &&	-144\\
$\bar{u}\bar{d}bb$	&&	$(1,0)$	&&	$BB$	&&	10604   &&  $a_1 = 1.7$, $a_2 = 3.2$, $a_3 = 4.9$	   &&	+51\\
$\bar{u}\bar{d}bb$	&&	$(1,1)$	&&	$BB^*$	&&	10663   &&  $a_1 = 1.6$, $a_2 = 4.1$, $a_3 = 4.4$	   &&	+55\\
$\bar{u}\bar{d}bb$	&&	$(1,2)$	&&	$B^*B^*$	&&	10695   &&  $a_1 = 1.9$, $a_2 = 12.1$, $a_3 = 3.8$	   &&	+31\\
$\bar{u}\bar{s}bb$	&&	$(1/2,0)$	&&	$B_s B$	&&	10679   &&  $a_1 = 2.2$, $a_2 = 4.1$, $a_3 = 6.5$	   &&	+57\\
$\bar{u}\bar{s}bb$	&&	$(1/2,1)$	&&	$B_s B^*$	&&	10635   &&  $a_1 = 3.7$, $a_2 = 16.2$, $a_3 = 5.6$	   &&	-42\\
$\bar{u}\bar{s}bb$	&&	$(1/2,2)$	&&	$B_s^* B^*$	&&	10781   &&  $a_1 = 2.2$, $a_2 = 7.4$, $a_3 = 5.5$	   &&	+45\\
$\bar{u}\bar{d}cb$	&&	$(0,0)$	&&	$DB$	&&	7124   &&  $a_1 = 2.7$, $a_2 = 6.3$, $a_3 = 5.3$	   &&	-17\\
$\bar{u}\bar{d}cb$	&&	$(0,1)$	&&	$DB^*$	&&	7179   &&  $a_1 = 2.7$, $a_2 = 6.3$, $a_3 = 4.8$	   &&	-18\\
$\bar{u}\bar{d}cb$	&&	$(0,2)$	&&	$D^*B^*$&&	7352   &&  $a_1 = 1.6$, $a_2 = 2.9$, $a_3 = 4.3$	   &&	+11\\
$\bar{u}\bar{s}cb$	&&	$(1/2,0)$	&&	$DB_s$	&&	7131   &&  $a_1 = 2.9$, $a_2 = 5.1$, $a_3 = 9.9$	   &&	-80\\
$\bar{u}\bar{s}cb$	&&	$(1/2,1)$	&&	$DB_s^*$	&&	7199   &&  $a_1 = 2.7$, $a_2 = 4.8$, $a_3 = 8.9$	   &&	-71\\
$\bar{u}\bar{s}cb$	&&	$(1/2,2)$	&&	$D^*B_s^*$	&&	7344   &&  $a_1 = 2.1$, $a_2 = 3.4$, $a_3 = 7.9$	   &&	-71\\
$\bar{u}\bar{s}cc$	&&	$(1/2,0)$	&&	$DD_s$	&&	3928   &&  $a_1 = 2.1$, $a_2 = 2.9$, $a_3 = 6.1$	   &&	+96\\
$\bar{u}\bar{s}cc$	&&	$(1/2,1)$	&&	$D^*D_s$	&&	4011   &&  $a_1 = 2.3$, $a_2 = 3.4$, $a_3 = 6.0$	   &&	+35\\
$\bar{u}\bar{s}cc$	&&	$(1/2,2)$	&&	$D^*D_s^*$	&&	4179   &&  $a_1 = 1.7$, $a_2 = 2.3$, $a_3 = 4.5$	   &&	+66\\
\hline 
\hline
\label{result}
\end{tabular}
\end{table}

\end{widetext}

To understand the origin of the strong attraction via color and CS interactions, it is crucial to investigate their effects on the binding energies of tetraquarks.
A complete understanding can be achieved only when a complete set is applied to calculating the mass of tetraquark systems.
For effective analysis, it is necessary to take into account the calculation of the mass of the $\bar{u}\bar{s}cb$ state using only the most dominant spatial basis.
For the purpose, we refit the mesons and baryons listed in Tables~\ref{mesons} and \ref{baryons} using the most dominant single spatial basis, as presented in Tables~\ref{Simplemesons} and \ref{Simplebaryons}.

The fitting results obtained using both a single spatial basis and a complete set indicate that the difference between the two schemes is at most 10 MeV, except for a few hadrons.
The model fitting parameters used to obtain the hadron masses presented in Tables~\ref{Simplemesons} and \ref{Simplebaryons} are as follows:
\begin{eqnarray}
&\kappa=107.7 \, \textrm{MeV fm}, \quad a_0=0.0327338 \, \textrm{(MeV$^{-1}$fm)$^{1/2}$},& \nonumber \\
&D=950  \, \textrm{MeV}, & \nonumber \\
&m_{u}=320 \, \textrm{MeV}, \qquad m_{s}=612 \, \textrm{MeV}, &\nonumber \\
&m_{c}=1893 \, \textrm{MeV}, \qquad m_{b}=5285 \, \textrm{MeV},	&\nonumber \\
&\alpha = 1.1349 \, \textrm{fm$^{-1}$}, \,\, \beta = 0.0011554 \, \textrm{(MeV fm)$^{-1}$}, &	\nonumber \\
&\gamma = 0.001370 \, \textrm{MeV$^{-1}$}, \,\, \kappa_0=230.244 \, \textrm{MeV}. 	 &
\label{SimpleFitParameters}
\end{eqnarray}

Table~\ref{Simpleresult} shows the masses of several tetraquarks obtained using the most dominant single spatial basis and the model fitting parameters given in Eq.~(\ref{SimpleFitParameters}).
It should be noted that, without employing the complete set, the $\bar{u}\bar{s}cb$ configurations in Table~\ref{result} cannot form bound states, showing a striking contrast compared to those in Table~\ref{Simpleresult}.

It is also interesting that the $\bar{u}\bar{d}cb$ with $(I,S) = (0,0)$ state is not as deeply bound as the $\bar{u}\bar{s}cb$ with $(I,S) = (1/2,0)$ state, despite both containing the same heavy quarks. This suggests a distinctive difference in the organization of stable tetraquark configurations. Therefore, we perform a comparative analysis of these two states.

\subsection{Comparison of the results for $\bar{u}\bar{s}cb$ between two calculation schemes}
\label{SecIIIA}
Theoretically, achieving an exact solution in our variational method requires fully applying the complete set.
Ensuring a sufficiently convergent solution in our calculations is great of importance within this framework.
Through consideration of up to the fifth quanta, it is observed that the ground state converges sufficiently within a few MeV\cite{Noh:2021lqs}.
Consequently, our study calculates the masses of tetraquarks while taking into account up to the fifth quanta.

\begin{table}[ht]

\caption{This table shows the masses of mesons obtained using only the most dominant single spatial basis and the parameter set given in Eq.~(\ref{SimpleFitParameters}).}	

\centering

\begin{tabular}{ccccc}
\hline
\hline	\multirow{2}{*}{Particle}	&	Experimental	&	Mass		&	Variational			& Error			\\
								&	Value (MeV)	&	(MeV)	&	Parameter (${\rm fm}^{-2}$)	&(\%)	\\
\hline 
$D$			&	1864.8		&	1860.4		&	$a$ = 4.4	&0.24	\\
$D^*$		&	2007.0		&	2005.2  		&	$a$ = 3.5	&0.08	\\
$\eta_{c}$	&	2983.6		&	2995.4		&	$a$ = 14.3	&0.39	\\
$J/\Psi$		&	3096.9		&	3117.5		&	$a$ = 11.0	&0.67	\\
$D_s$		&	1968.3		&	1961.0		&	$a$ = 7.0	&0.37	\\
$D^*_s$		&	2112.1		&	2097.6		&	$a$ = 5.4	&0.69	\\
$K$			&	493.68		&	503.7		&	$a$ = 14.3	&1.23	\\
$K^*$		&	891.66	 	&	875.0		&	$a$ = 2.6	&1.87	\\

$B$			&	5279.3	 	&	5283.5		&	$a$ = 4.3	&0.07	\\
$B^*$		&	5325.2		&	5339.2		&	$a$ = 3.9	&0.27	\\
$\eta_b$		&	9398.0		&	9370.6		&	$a$ = 52.7	&0.30	\\
$\Upsilon$	&	9460.3		&	9485.7		&	$a$ = 40.8	&0.27	\\
$B_s$		&	5366.8		&	5351.2		&	$a$ = 7.4	&0.29	\\
$B_s^*$		&	5415.4		&	5410.3		&	$a$ = 6.6	&0.09	\\
$B_c$		&	6275.6		&	6273.2		&	$a$ = 21.2	&0.02	\\
$B_c^*$		&		...		&	6357.7		&	$a$ = 17.4	&...	\\
\hline 
\hline
\label{Simplemesons}
\end{tabular}
\end{table}
\begin{table}[ht]
\caption{This table shows the masses of baryons obtained using only the most dominant single spatial basis and the parameter set given in Eq.~(\ref{SimpleFitParameters}).}

\centering

\begin{tabular}{ccccc}
\hline
\hline	\multirow{2}{*}{Particle}	&	Experimental	&	Mass		&	\quad Variational\quad			& Error			\\
								&	Value (MeV)	&	(MeV)	&	\quad Parameters (${\rm fm}^{-2}$)\quad	& (\%)\\
\hline  
$\Lambda$		&	1115.7	&	1112.0	&	\quad$a_1$ = 2.6, $a_2$ = 2.6\quad	&0.33	\\
$\Lambda_{c}$	&	2286.5	&	2265.4	&	\quad$a_1$ = 2.7, $a_2$ = 3.5\quad	&0.92	\\
$\Xi_{cc}$		&	3621.4	&	3609.6	&	\quad$a_1$ = 7.3, $a_2$ = 3.0\quad	&0.33	\\
$\Lambda_b$		&	5619.4	&	5608.8	&	\quad$a_1$ = 2.7, $a_2$ = 3.9\quad	&0.19	\\
$\Sigma_{c}$		&	2452.9	&	2444.6	&	\quad$a_1$ = 2.0, $a_2$ = 3.5\quad	&0.37	\\
$\Sigma_{c}^*$	&	2517.5	&	2528.2	&	\quad$a_1$ = 1.8, $a_2$ = 3.1\quad	&0.39	\\
$\Sigma_{b}$		&	5811.3	&	5825.5	&	\quad$a_1$ = 1.9, $a_2$ = 3.8\quad	&0.17	\\
$\Sigma_{b}^*$	&	5832.1	&	5858.2	&	\quad$a_1$ = 1.9, $a_2$ = 3.6\quad	&0.40	\\
$\Sigma$			&	1192.6	&	1199.7	&	\quad$a_1$ = 2.0, $a_2$ = 2.9\quad	&0.59	\\
$\Sigma^*$		&	1383.7	&	1405.3	&	\quad$a_1$ = 1.7, $a_2$ = 2.2\quad	&1.56	\\
$\Xi$			&	1314.9	&	1316.6	&	\quad$a_1$ = 3.2, $a_2$ = 2.7\quad	&0.13	\\
$\Xi^*$			&	1531.8	&	1532.7	&	\quad$a_1$ = 2.8, $a_2$ = 2.1\quad	&0.06	\\

$\Xi_{c}$		&	2467.8	&	2459.5	&	\quad$a_1$ = 3.1, $a_2$ = 4.4\quad	&0.44	\\
$\Xi_{c}^*$		&	2645.9	&	2639.9	&	\quad$a_1$ = 2.3, $a_2$ = 4.0\quad	&0.24	\\
$\Xi_{b}$		&	5787.8	&	5790.3	&	\quad$a_1$ = 3.1, $a_2$ = 5.2\quad	&0.03	\\
$\Xi_{b}^*$		&	5945.5	&	5958.0	&	\quad$a_1$ = 2.4, $a_2$ = 4.9\quad	&0.04	\\

$p$				&	938.27	&	958.93	&	\quad$a_1$ = 2.2, $a_2$ = 2.2\quad	&2.20	\\
$\Delta$			&	1232		&	1269.5	&	\quad$a_1$ = 1.7, $a_2$ = 1.7\quad	&3.04	\\
\hline 
\hline
\label{Simplebaryons}
\end{tabular}
\end{table}

The masses derived from such calculations significantly differ from those obtained using only the most dominant single spatial basis. Table~\ref{Simpleresult} shows that even the deeply bound $\bar{u}\bar{s}cb$ with $(I,S) = (1/2,0)$ state in Table~\ref{result} becomes unbound when using only the most dominant single spatial basis.
This substantial difference originates from wave functions other than the most dominant single spatial wave function, suggesting an influence from internal states.

To investigate the impact of internal states on the binding energy of the $\bar{u}\bar{s}cb$, it is crucial to compare results from two calculation schemes, as shown in Table~\ref{TetraSeparation}: Scheme 1 represents results obtained using only the most dominant single spatial basis, while Scheme 2 represents results using a complete set up to the fifth quanta.
Furthermore, for the $\bar{u}\bar{s}cb$, its thresholds between the two schemes are fitted almost equally, with values of 7210.6 MeV and 7211.6 MeV, respectively. Therefore, by comparing the contributions of each term to the mass of the tetraquark, it is possible to analyze where the enhanced attraction arises in Scheme 2.
For this comparison, we will analyze the contributions from each part of the Hamiltonian to the mass of the $\bar{u}\bar{s}cb$.
Table~\ref{TetraSeparation} presents these contributions for the $\bar{u}\bar{s}cb$ with $(I,S) = (1/2,0)$ state under both calculation schemes.

We begin by comparing the kinetic energy part.
Since the kinetic energy is inversely proportional to the sizes of quark pairs\cite{Park:2018wjk},
the kinetic energy in Scheme 2 is smaller than that in Scheme 1. The relevant sizes for the kinetic energy are those of the $(1,2)$ and $(3,4)$ pairs, and the distance between the $(1,2)$ and $(3,4)$ pairs, as shown in Table~\ref{RelDistances}.
From Table~\ref{TetraSeparation}, it can be shown that the kinetic energy part makes a significant contribution, second only to the confinement potential part, which will be discussed subsequently.
Although the fitted quark masses in Schemes 1 and 2 are very close to each other, differences in the sizes of each quark pair are responsible for the difference in kinetic energy.
Table~\ref{RelDistances} shows that the sizes of the $(1,2)$, $(3,4)$ quark pairs, as well as the distance between the quark pairs $(1,2)$ and $(3,4)$, are larger in Scheme 2.
Consequently, the kinetic energy is lower in Scheme 2.

We now discuss the confinement potential $V^C$ part presented in Table~\ref{TetraSeparation}.
As indicated in Eq.~(\ref{ConfineP}), the value of the confinement potential is determined by the matrix element, $\lambda^{c}_{i} \lambda^{c}_{j}$ as well as the spatial functional form.
To quantitatively investigate the color interaction through our approach, we introduce the factor related to the color interaction, $-\langle \lambda^{c}_{i} \lambda^{c}_{j} \rangle$, which is calculated with respect to the ground state.
Also, in this approach, the expectation value of the confinement potential, $\langle V^C_{ij} \rangle$, can be factorized into two terms in Scheme 1, as shown in Eq.~(\ref{ExpecVC-1G}): it consists of the color interaction part and the spatial part.
\begin{eqnarray}
\langle V^C_{ij} \rangle
=
-\frac{3}{16}\langle \lambda^{c}_{i} \lambda^{c}_{j} \rangle \left\langle - \frac{\kappa}{r_{ij}} + \frac{r_{ij}}{a^2_0} \right\rangle.
\label{ExpecVC-1G}
\end{eqnarray}
This is evident because a single spatial function is used in Scheme 1 calculations, as will be discussed later.
We quantitatively analyze our numerical results using the color interaction factor and the spatial factor for Scheme 1.
In particular, this analysis of the spatial part becomes possible from the manifest correlation between the magnitude of the potential and the size.
Consequently, in comparing the two calculation schemes, this analysis is most effective in assessing the impact of employing the complete set of harmonic oscillator bases.

\begin{widetext}

\begin{table}[h]
\caption{This table represents the results obtained using only the most dominant single spatial basis and the model parameters given in Eq.~(\ref{SimpleFitParameters}).}
\begin{tabular}{ccccccccccc}
\hline
\hline
Configuration		&&	$(I,S)$	&&	Thresholds	&&	Mass (MeV)	&&  Variational Parameters (fm$^{-2}$)    &&  $B_{T}$ (MeV)	\\
\hline 
$\bar{u}\bar{s}bb$	&&	$(1/2,1)$	&&	$B_s B^*$	&&	10675   &&  $a_1 = 3.1$, $a_2 = 18.7$, $a_3 = 3.6$	   &&	-16\\
$\bar{u}\bar{d}cb$	&&	$(0,0)$	&&	$DB$	&&	7207   &&  $a_1 = 2.7$, $a_2 = 10.8$, $a_3 = 3.1$	   &&	+52\\
$\bar{u}\bar{d}cb$	&&	$(0,1)$	&&	$DB^*$	&&	7243   &&  $a_1 = 2.7$, $a_2 = 10.0$, $a_3 = 2.9$	   &&	+43\\
$\bar{u}\bar{s}cb$	&&	$(1/2,0)$	&&	$DB_s$	&&	7297   &&  $a_1 = 3.0$, $a_2 = 8.4$, $a_3 = 7.4$	   &&	+85\\
$\bar{u}\bar{s}cb$	&&	$(1/2,1)$	&&	$DB_s^*$	&&	7357   &&  $a_1 = 3.1$, $a_2 = 10.1$, $a_3 = 5.1$	   &&	+86\\
$\bar{u}\bar{s}cb$	&&	$(1/2,2)$	&&	$D^*B_s^*$	&&	7512   &&  $a_1 = 2.3$, $a_2 = 5.2$, $a_3 = 7.1$	   &&	+96\\
\hline 
\hline
\label{Simpleresult}
\end{tabular}
\end{table}
\begin{table}[h]

\caption{Contributions from each part of the Hamiltonian to the mass of the $\bar{u}\bar{s}cb$ with $(I,S) = (1/2,0)$ state. Here, $(i,j)$ denotes the quark pair in the $\bar{u}\bar{s}cb$ configuration. The term $D$ is excluded from $V^C(i,j)$. All values are given in MeV.}
\centering

\begin{tabular}{ccccccccccccc}
\hline
\hline
\multirow{2}{*}{Type}&& \multirow{2}{*}{$\sum^{4}_{i=1} m_i$}&\multirow{2}{*}{$\sum^{3}_{i=1} \frac{{\mathbf p}^{2}_{\mathbf{x}_i}}{2 m'_i}$}&\multirow{2}{*}{Subtotal}&& \multicolumn{6}{c}{$V^{CS}$}\\
\cline{7-13}
&&&&&&(1,2)&(1,3)&(1,4)&(2,3)&(2,4)&(3,4)&Subtotal\\
\hline
Scheme 1    && 	8110.0	&	927.4    &   9037.4 &&  -65.40   &   -67.28  &    -16.58   &   -35.38   &   -33.00   &   -8.643   &   -226.3 \\
Scheme 2    && 	8094.0	&	866.5    &   8960.5 &&  -22.51   &   -92.06  &    -11.98   &   -29.10   &   -39.97   &   -2.195   &   -197.8  \\
\hline
\hline
\multirow{2}{*}{Type}&& \multicolumn{6}{c}{$V^{C}$}    &&&  \multirow{2}{*}{$-2D$} &&  \multirow{2}{*}{Total}\\
\cline{3-9}
&&(1,2)&(1,3)&(1,4)&(2,3)&(2,4)&(3,4)&Subtotal\\
\hline
Scheme 1    && 	38.78   &   211.6  &    69.24   &   45.25   &   19.76   &   1.086   &   385.7  &&    -1900.0		&&   7296.8\\
Scheme 2    && 	-10.65  &   204.4  &    61.61   &   22.51   &   22.44   &   -14.28  &   286.0 &&    -1918.0		&&   7130.7\\
\hline
\hline
\label{TetraSeparation}
\end{tabular}
\end{table}

\end{widetext}

Before analyzing the numerical results of the $V^C$ part in Table~\ref{TetraSeparation} in detail, it is necessary to examine the value of the spatial part of the confinement potential when the size of a quark pair is small enough.
From Eq.~(\ref{ExpecVC-1G}), it is roughly estimated that the spatial part goes to zero near 0.32 fm.
Thus, if the size of a quark pair is comparable to 0.32 fm, it results in minimal contributions to the confinement part, as indicated in Table~\ref{TetraSeparation}. In addition, it should be noted that the attraction in each confinement part is primarily attributed to the color interaction factor, $-\langle \lambda^{c}_{i} \lambda^{c}_{j} \rangle$, in Eq.~(\ref{ExpecVC-1G}).

On the other hand, when a complete set is introduced in the calculations (Scheme 2), the expectation value, $\langle V^C_{ij} \rangle$, is decomposed into two parts. One has a form of Eq.~(\ref{ExpecVC-1G}), and the other part comes from excited states.
\begin{eqnarray}
\hspace{-0.8cm}\left\langle V^C_{ij} \right\rangle
&=&
\left\langle \Psi_1 \left|  -\lambda^{c}_{i} \lambda^{c}_{j} \, f(r_{ij}) \right| \Psi_1 \right\rangle
\nonumber \\
\hspace{-0.8cm}&=&
\sum^{N_{\rm max}}_{n=1} \left\langle \Psi_1 \left|  -\lambda^{c}_{i} \lambda^{c}_{j} \right| \Psi_n \right\rangle \left\langle \Psi_n \left|  f(r_{ij}) \right| \Psi_1 \right\rangle
\nonumber \\
\hspace{-0.8cm}&=&
\left\langle \Psi_1 \left|  -\lambda^{c}_{i} \lambda^{c}_{j} \right| \Psi_1 \right\rangle \left\langle \Psi_1 \left|  f(r_{ij}) \right| \Psi_1 \right\rangle
\nonumber \\
\hspace{-0.8cm}
&&+ \sum^{N_{\rm max}}_{n=2} \left\langle \Psi_1 \left|  -\lambda^{c}_{i} \lambda^{c}_{j} \right| \Psi_n \right\rangle \left\langle \Psi_n \left|  f(r_{ij}) \right| \Psi_1 \right\rangle,
\label{VcExpansion}
\end{eqnarray}
where $N_{\rm max}$ is the total number of eigenstates of the Hamiltonian and $f(r_{ij})$ is the spatial part of $V^C$ as follows.
\begin{eqnarray}
f(r_{ij}) = - \frac{\kappa}{r_{ij}} + \frac{r_{ij}}{a^2_0}.
\end{eqnarray}
Here, $\left| \Psi_1 \right\rangle$ is the ground state(lowest eigenstate) and $\left| \Psi_n \right\rangle$ is the $n$th eigenstate of the Hamiltonian, as shown below:
\begin{eqnarray}
\hspace{-0.5cm}\left| \Psi_n \right\rangle
&=&
\sum_k^{k_{\rm max}} [ C_1 (n,k) |1 \rangle_{CS} + C_2 (n,k) |2 \rangle_{CS}
\nonumber \\
\hspace{-0.5cm}&&\hspace{0.5cm}+ C_3 (n,k) |3 \rangle_{CS} + C_4 (n,k) |4 \rangle_{CS} ] |\phi_k\rangle,
\label{H_eigenstate}
\end{eqnarray}
where $|\phi_k\rangle$ is the $k$th harmonic oscillator basis and $k_{\rm max}$ is the total number of harmonic oscillator bases representing each eigenstate of the Hamiltonian. For each $\left| \Psi_n \right\rangle$, $C_l(n,k)$ gives the probability amplitude for the wave function composed of the $l$th CS basis and the $k$th spatial basis, $|\phi_k\rangle$.
The four CS bases are as follows.
\begin{eqnarray}
\hspace{-0.6cm}&|1\rangle_{CS} \equiv | \{\bar{q}_1\bar{q}_2\}^{\mathbf{3}}_1 \{Q_3 Q_4\}^{\mathbf{\bar{3}}}_1 \rangle, \,\, |2\rangle_{CS} \equiv | \{\bar{q}_1\bar{q}_2\}^{\mathbf{3}}_0 \{Q_3 Q_4\}^{\mathbf{\bar{3}}}_0 \rangle,&
\nonumber \\
\hspace{-0.6cm}&|3\rangle_{CS} \equiv | \{\bar{q}_1\bar{q}_2\}^{\mathbf{\bar{6}}}_1 \{Q_3 Q_4\}^{\mathbf{6}}_1 \rangle, \,\, |4\rangle_{CS} \equiv | \{\bar{q}_1\bar{q}_2\}^{\mathbf{\bar{6}}}_0 \{Q_3 Q_4\}^{\mathbf{6}}_0 \rangle.&
\nonumber \\
\label{Tetra-CS-bases}
\end{eqnarray}
Here, in the case of the $\bar{u}\bar{s}cb$, $\bar{q}_1= \bar{u}$, $\bar{q}_2= \bar{s}$, $Q_3=c$, and $Q_4=b$.
Note that these four CS bases are orthogonal to each other.

\begin{table}[h]

\caption{Relative distances between the quarks in the $\bar{u}\bar{s}cb$ with $(I,S) = (1/2,0)$ state. In the table, $(i,j)$ denotes the quark pair and $(1,2)$-$(3,4)$ the distance between the centers of masses of the quark pairs $(1,2)$ and $(3,4)$. All distances are provided in fm unit.}	
\centering
\begin{tabular}{ccccccccc}
\hline
\hline
Quark Pair	&$(1,2)$&$(1,3)$&$(1,4)$&$(2,3)$&$(2,4)$&	$(3,4)$&$(1,2)$-$(3,4)$\\
\hline
Scheme 1		&0.651	&0.604	&0.542	&0.482	&0.401	&0.389	&0.316\\
Scheme 2		&0.855	&0.585	&0.736	&0.673	&0.440	&0.655	&0.322\\
\hline 
\hline
\label{RelDistances}
\end{tabular}
\end{table}

\begin{table}[h]

\caption{Color interaction factor, $-\langle \lambda^{c}_{i} \lambda^{c}_{j} \rangle$, and color-spin(CS) interaction factor, $-\langle \lambda^{c}_{i} \lambda^{c}_{j} \boldsymbol{\sigma}_i \cdot \boldsymbol{\sigma}_j \rangle$, for each quark pair $(i,j)$ in the $\bar{u}\bar{s}cb$ with $(I,S)=(1/2,0)$ state.}	
\centering
\begin{tabular}{ccccccc}
\hline
\hline
\multirow{2}{*}{Quark pair}	&& \multicolumn{2}{c}{$-\langle \lambda^{c}_{i} \lambda^{c}_{j} \rangle$} && \multicolumn{2}{c}{$-\langle \lambda^{c}_{i} \lambda^{c}_{j} \boldsymbol{\sigma}_i \cdot \boldsymbol{\sigma}_j \rangle$}	\\
\cline{3-4}\cline{6-7}
        && Scheme 1&Scheme 2	&&	Scheme 1&	Scheme 2	\\
\hline
$(1,2)$ &&	0.520  &  -0.074    &&	-4.176  &  -1.493        \\
$(1,3)$ &&	3.350  &  4.114     &&	-10.81  &  -12.59        \\
$(1,4)$ &&	1.463  &  1.294     &&	-6.047  &  -4.347     \\
$(2,3)$ &&	1.463  &  1.294     &&	-6.047  &  -4.347     \\
$(2,4)$ &&	3.350  &  4.114     &&	-10.81  &  -12.59     \\
$(3,4)$ &&	0.520  &  -0.074    &&	-4.176  &  -1.493     \\
\hline 
\hline
\label{C-CS-Factors-uscb}
\end{tabular}
\end{table}

They are associated with each harmonic oscillator basis in constructing the total wave function.
For instance, consider the $\bar{u}\bar{d}cb$ with $(I,S)=(0,0)$ state.
Here, the Pauli principle requires that the CS bases associated with $\psi^{spatial}_{[0,0,0,0,0,0]}$ be symmetric under the permutation $(12)$, since $\psi^{spatial}_{[0,0,0,0,0,0]}$ in Eq.~(\ref{Spatial_Wave}) is symmetric under the permutation $(12)$.
Therefore, there are two CS bases corresponding to $\psi^{spatial}_{[0,0,0,0,0,0]}$. Conversely, for $\psi^{spatial}_{[0,0,0,1,1,0]}$, the corresponding CS basis should be antisymmetric under the permutation $(12)$ since $\psi^{spatial}_{[0,0,0,1,1,0]}$ in Eq.~(\ref{Spatial_Wave}) is antisymmetric under the permutation $(12)$.
One can check the symmetry of the spatial part of the total wave function from Eq.~(\ref{CoordSet1}) and Eq.~(\ref{Spatial_Wave}).
Consequently, only two of the four CS bases correspond to each spatial basis for the $\bar{u}\bar{d}cb$ with $(I,S)=(0,0)$ state due to the symmetry of the light quark sector.

In contrast, for the $\bar{u}\bar{s}cb$ with $(I,S)=(0,0)$ state, all four CS bases are associated with each individual spatial basis because there is no symmetry constraints imposed by the Pauli principle.
Therefore, in Eq.~(\ref{VcExpansion}), $N_{\rm max}$ equals $4 k_{\rm max}$ for the $\bar{u}\bar{s}cb$ state when computing the mass.

On the other hand, in Scheme 1, there is only one spatial basis($k_{\rm max}=1$) and thus the term $\left\langle \Psi_n \left| f(r_{ij}) \right| \Psi_1 \right\rangle$ in the second term in Eq.~(\ref{VcExpansion}) is rewritten as follows.
\begin{eqnarray}
\hspace{-0.0cm}&&\left\langle \Psi_n \left| f(r_{ij}) \right| \Psi_1 \right\rangle
\nonumber \\
\hspace{-0.0cm}&&=
[C_1(n,1) C_1(1,1) + C_2(n,1) C_2(1,1) + C_3(n,1) C_3(1,1)
\nonumber \\
\hspace{-0.0cm}&&\hspace{0.4cm}+ C_4(n,1) C_4(1,1)] \langle \phi_1| f(r_{ij}) |\phi_1\rangle,
\label{Scheme1_Eigen}
\end{eqnarray}
where $n=2,3,4$.
In this case, the eigenstates are simply given as follows.
\begin{eqnarray}
\hspace{-0.5cm}\left| \Psi_n \right\rangle
&=&
[ C_1 (n,1) |1 \rangle_{CS} + C_2 (n,1) |2 \rangle_{CS}
\nonumber \\
\hspace{-0.5cm}&&\hspace{0.5cm}+ C_3 (n,1) |3 \rangle_{CS} + C_4 (n,1) |4 \rangle_{CS} ] |\phi_1\rangle ,
\nonumber \\
\end{eqnarray}
where $n=1,2,3,4$.
Then, the orthogonality of these eigenstates ensures that Eq.~(\ref{Scheme1_Eigen}) be equal to zero in Scheme 1.
In contrast, in Scheme 2, Eq.~(\ref{Scheme1_Eigen}) becomes more complicated, reflecting the influence of the internal states as follows:
\begin{eqnarray}
\hspace{-0.0cm}&&\left\langle \Psi_n \left| f(r_{ij}) \right| \Psi_1 \right\rangle
=
\sum_{k,l} [C_1(n,k) C_1(1,l) + C_2(n,k) C_2(1,l)
\nonumber \\
\hspace{-2.5cm}&&\hspace{0.0cm} + C_3(n,k) C_3(1,l) + C_4(n,k) C_4(1,l)] \langle \phi_k| f(r_{ij}) |\phi_l\rangle,
\label{VcExpansion2}
\end{eqnarray}
where $n=2,3,4,\dots,N_{\rm max}$. Therefore, Eq.~(\ref{VcExpansion2}) cannot simply vanish because more than one spatial function is involved.
In this respect, we designate the first term in Eq.~(\ref{VcExpansion}) as the `non-mixing' term and the second term as `mixing' term.

We are now prepared to investigate the fundamental structure of the tetraquark states by comparing the two calculation schemes.
For the $\bar{u}\bar{s}cb$ configuration, this comparison involves analyzing the sizes of quark pairs from Table~\ref{RelDistances}, the color interaction factors from Table~\ref{C-CS-Factors-uscb}, and the detailed contributions of $V^C$ part from Table~\ref{Detailed_Decomp_uscb}.

Since a quantitative analysis can only be made through the non-mixing term, we separate the contributions of $V^C$ into two parts: the non-mixing and mixing terms as detailed in Table~\ref{Detailed_Decomp_uscb}.
Subsequently, we specifically compare these numerical results by incorporating the the quark pair sizes from Table~\ref{RelDistances} and the color interaction factors from Table~\ref{C-CS-Factors-uscb}.
The most important effect of introducing the complete set becomes apparent through the contributions of the mixing terms in Table~\ref{Detailed_Decomp_uscb}, which are entirely absent in Scheme 1 calculations.
The subtotal values in Table~\ref{Detailed_Decomp_uscb} correspond to the values in Table~\ref{TetraSeparation}.

As shown in Table~\ref{TetraSeparation}, the total value of non-mixing term in Scheme 2 is significantly greater than that in Scheme 1.
However, the contribution of the confinement $V^C$ part in Scheme 2 shows a notably reduced repulsive effect compared to Scheme 1, as indicated in Table~\ref{TetraSeparation}.
This shift towards attraction is attributed to the effect of mixing terms resulting from the introduction of the complete set.

\begin{table}[ht]

\caption{``Non-mixing'' and ``Mixing'' components of $V^C$ part presented in Table~\ref{TetraSeparation}. All values are provided in MeV unit.}	
\centering
\begin{tabular}{cccccc}
\hline
\hline
\multirow{2}{*}{Quark Pair} &&&	\multicolumn{3}{c}{Scheme 1}	\\
\cline{4-6}
&&&Non-mixing&Mixing&Subtotal\\
\hline
    	$(1,2)$			&&&	38.78    &  0.0   &   38.78       \\
    	$(1,3)$			&&&	211.6    &	0.0   &   211.6       \\
    	$(1,4)$			&&&	69.24    &	0.0   &   69.24       \\
    	$(2,3)$			&&&	45.25    &	0.0   &   45.25       \\
    	$(2,4)$			&&&	19.76    &	0.0   &   19.76       \\
        $(3,4)$			&&&	1.086    &	0.0   &   1.086       \\
\hline
        Total			&&&	385.7    &	0.0   &   385.7       \\
\hline 
\hline
\multirow{2}{*}{Quark Pair} &&&	\multicolumn{3}{c}{Scheme 2}      \\
\cline{4-6}
&&&Non-mixing&Mixing&Subtotal\\
\hline
    	$(1,2)$			&&&	-9.042   &  -1.610   &   -10.65       \\
    	$(1,3)$			&&&	248.7    &	-44.30    &   204.4       \\
    	$(1,4)$			&&&	123.8    &	-62.16   &   61.61       \\
    	$(2,3)$			&&&	103.9    &	-81.40   &   22.51       \\
    	$(2,4)$			&&&	80.51    &	-58.07    &   22.44       \\
        $(3,4)$			&&&	-5.822   &	-8.460   &   -14.28       \\
\hline
        Total			&&&	542.0    &	-256.0   &   286.0       \\
\hline  
\hline
\label{Detailed_Decomp_uscb}
\end{tabular}
\end{table}

Let us first quantitatively analyze the confinement $V^C$ part.
For the $(1,2)$ pair, although the size in Scheme 2 is larger than that in Scheme 1, the magnitude of the non-mixing component in Scheme 2 is remarkably smaller, and the sign is reversed compared to that in Scheme 1.
This is because $-\langle \lambda^{c}_{i} \lambda^{c}_{j} \rangle$ is smaller by an order of magnitude and the sign is reversed in Scheme 2, as shown in Table~\ref{C-CS-Factors-uscb}.
However, for the $(3,4)$ pair, even though $-\langle \lambda^{c}_{i} \lambda^{c}_{j} \rangle$ is the same as that of the $(1,2)$ pair and the size is also larger in Scheme 2, the magnitude of the non-mixing component of the $(3,4)$ pair in Table~\ref{Detailed_Decomp_uscb} increases in Scheme 2.
The size of the $(3,4)$ pair in Scheme 1 is close to 0.32 fm, where the spatial part of $V^C$, as shown in Eq.~(\ref{ExpecVC-1G}), nearly goes to zero.
Therefore, the magnitude of the $(3,4)$ pair in Scheme 1 becomes considerably smaller, regardless of the corresponding value of $-\langle \lambda^{c}_{i} \lambda^{c}_{j} \rangle$.

For the $(1,3)$ pair, the magnitude of the non-mixing component in Scheme 2 is initially expected to be slightly smaller due to the reduced size in Scheme 2.
However, in this case, the difference in the magnitude of the non-mixing component between the two schemes is decisively influenced by the magnitude of $-\langle \lambda^{c}_{i} \lambda^{c}_{j} \rangle$.

For the $(2,4)$ pair, both the magnitude of $-\langle \lambda^{c}_{i} \lambda^{c}_{j} \rangle$ and the size are larger in Scheme 2.
Moreover, the size change occurs within a range much closer to 0.32 fm compared to the $(1,3)$ pair.
When the size of a quark pair is approximately 0.32 fm, the spatial part of the confinement potential, $V^C$, behaves in a non-linear manner with respect to changes in the size of the quark pair. 
Therefore, the difference in the non-mixing component between the two schemes is much larger for the $(2,4)$ pair than for the $(1,3)$ pair.

For the $(1,4)$ and $(2,3)$ pairs, the non-mixing components increase significantly in Scheme 2.
This increase is primarily due to the larger sizes of both the $(1,4)$ and $(2,3)$ pairs in Scheme 2.
Furthermore, since the values of $-\langle \lambda^{c}_{i} \lambda^{c}_{j} \rangle$ do not differ significantly between the two schemes, this does not counteract the trend toward the increase of the non-mixing component.

\begin{table}[ht]

\caption{Simliar to Table~\ref{Detailed_Decomp_uscb} but for the $V^{CS}$ part presented in Table~\ref{TetraSeparation}.}	
\centering
\begin{tabular}{cccccc}
\hline
\hline
\multirow{2}{*}{Quark Pair} &&&	\multicolumn{3}{c}{Scheme 1}	\\
\cline{4-6}
&&&Non-mixing&Mixing&Subtotal\\
\hline
    	$(1,2)$			&&&	-65.40    & 0.0   &   -65.40       \\
    	$(1,3)$			&&&	-67.28    &	0.0   &   -67.28       \\
    	$(1,4)$			&&&	-16.58    &	0.0   &   -16.58       \\
    	$(2,3)$			&&&	-35.38    &	0.0   &   -35.38       \\
    	$(2,4)$			&&&	-33.00    &	0.0   &   -33.00       \\
            $(3,4)$			&&&	-8.643    &	0.0   &   -8.643       \\
\hline 
\hline
\multirow{2}{*}{Quark Pair} &&&	\multicolumn{3}{c}{Scheme 2}      \\
\cline{4-6}
&&&Non-mixing&Mixing&Subtotal\\
\hline
    	$(1,2)$			&&&	-14.54   &  -7.962    &   -22.51       \\
    	$(1,3)$			&&&	-83.81   &	-8.242    &   -92.06       \\
    	$(1,4)$			&&&	-7.494   &	-4.489    &   -11.98       \\
    	$(2,3)$			&&&	-16.03   &	-13.07    &   -29.10       \\
    	$(2,4)$			&&&	-35.39   &	-4.581    &   -39.97       \\
    $(3,4)$			&&&	-1.153   &	-1.042   &   -2.195       \\
\hline  
\hline
\label{Detailed_CS_Decomp_uscb}
\end{tabular}
\end{table}

Similar to the discussion of the $V^C$ part, we also separate the contributions from the $V^{CS}$ part into non-mixing and mixing terms, as presented in Table~\ref{Detailed_CS_Decomp_uscb}.
As shown in Eq.~(\ref{CSP}), the value of the hyperfine potential depends on the matrix element, $\lambda^{c}_{i} \lambda^{c}_{j} \boldsymbol{\sigma}_i \cdot \boldsymbol{\sigma}_j$, as well as the spatial functional form.
To quantitatively investigate the hyperfine potential part, we introduce the CS interaction factor, $-\langle \lambda^{c}_{i} \lambda^{c}_{j} \boldsymbol{\sigma}_i \cdot \boldsymbol{\sigma}_j \rangle$, calculated as the expectation values of the CS matrices.
We then compare the numerical results presented in Table~\ref{Detailed_CS_Decomp_uscb} by incorporating the sizes of quark pairs from Table~\ref{RelDistances} and the CS interaction factors, $-\langle \lambda^{c}_{i} \lambda^{c}_{j} \boldsymbol{\sigma}_i \cdot \boldsymbol{\sigma}_j \rangle$, from Table~\ref{C-CS-Factors-uscb}.

For the $(1,2)$ and $(3,4)$ pairs, the magnitudes of the non-mixing components decrease in Scheme 2.
This reduction is attributed to the smaller magnitude of $-\langle \lambda^{c}_{i} \lambda^{c}_{j} \boldsymbol{\sigma}_i \cdot \boldsymbol{\sigma}_j \rangle$ in Scheme 2, as well as the larger sizes of both the $(1,2)$ and $(3,4)$ pairs compared to those in Scheme 1.
On the other hand, there is another point we should remark is that the reduction ratio of the non-mixing component in Scheme 2 is more pronounced for the $(3,4)$ pair than for the $(1,2)$ pair.
This difference arises from the spatial functional form of the hyperfine potential, which resembles a Coulomb type.
Furthermore, the size change of the $(3,4)$ pair occurs within a shorter range.

For the $(1,3)$ pair, the magnitude of the non-mixing term in Scheme 2 surpasses that in Scheme 1 due to its smaller size and the larger magnitude of $-\langle \lambda^{c}_{i} \lambda^{c}_{j} \boldsymbol{\sigma}_i \cdot \boldsymbol{\sigma}_j \rangle$ in Scheme 2.

For the $(2,4)$ pair, the magnitude of the non-mixing component in Scheme 2 slightly increases, despite the larger size compared to Scheme 1 and the same ratio of $-\langle \lambda^{c}_{i} \lambda^{c}_{j} \boldsymbol{\sigma}_i \cdot \boldsymbol{\sigma}_j \rangle$ as in the case of the $(1,3)$ pair.
From Table~\ref{Detailed_CS_Decomp_uscb}, the ratio of the magnitude of the non-mixing term in Scheme 2 to that in Scheme 1 is calculated as 1.07.
This value is derived roughly from the ratio of the inverse sizes and the ratio of the values of $-\langle \lambda^{c}_{i} \lambda^{c}_{j} \boldsymbol{\sigma}_i \cdot \boldsymbol{\sigma}_j \rangle$.

For the $(1,4)$ and $(2,3)$ pairs, the magnitudes of the non-mixing terms in Scheme 2 are smaller compared to Scheme 1.
This reduction is primarily due to the larger sizes of both quark pairs in Scheme 2. Additionally, the smaller magnitude of $-\langle \lambda^{c}_{i} \lambda^{c}_{j} \boldsymbol{\sigma}_i \cdot \boldsymbol{\sigma}_j \rangle$ in Scheme 2 further contributes to this observation.

Based on our quantitative analysis, we find that for the deeply bound state of $\bar{u}\bar{s}cb$, the significant enhancement in attraction compared to Scheme 1 primarily arises from the $V^C$ part, as indicated in Table~\ref{TetraSeparation}.
Indeed, the presence of the mixing term in Scheme 2 calculation is the very important factor that allows the $\bar{u}\bar{s}cb$ state to become deeply bound.

\subsection{Principal differences between $\bar{u}\bar{d}cb$ and $\bar{u}\bar{s}cb$ states}

The variation in isospin configurations among tetraquarks containing two heavy quarks, $c$ and $b$, results in significant differences in binding energy, as shown in Table~\ref{result}.
To understand these distinctive differences between them, we proceed with a comparative analysis of both the $\bar{u}\bar{d}cb$ and $\bar{u}\bar{s}cb$ with spin 0 states, following the methodmmmmmm performed in the previous section.

For this purpose, as presented in Table~\ref{TetraSeparation1}, we analyze the contributions of $V^{C}$ and $V^{CS}$ for each quark pair in both the $\bar{u}\bar{d}cb$ and $\bar{u}\bar{s}cb$ states. We calculate the sizes of the quark pairs and evaluate the color and CS interaction factors, $-\langle \lambda^{c}_{i} \lambda^{c}_{j} \rangle$ and $-\langle \lambda^{c}_{i} \lambda^{c}_{j} \boldsymbol{\sigma}_i \cdot \boldsymbol{\sigma}_j \rangle$, for both configurations, as presented in Table~\ref{C-CS-Factors-udcb-uscb}.

\begin{table}[h]

\caption{Contributions from the confinement potential $V^{C}$ and hyperfine potential $V^{CS}$ for the $\bar{u}\bar{d}cb$ with $(I,S) = (0,0)$ and the $\bar{u}\bar{s}cb$ with $(I,S) = (1/2,0)$ states. The values for $V^{C}$ and $V^{CS}$ are expressed in MeV unit. The quark pair length $l$ is provided in fm unit.}
\centering

\begin{tabular}{ccccccccc}
\hline
\hline
\multirow{2}{*}{Quark pair}	&& \multicolumn{3}{c}{$\bar{u}\bar{d}cb$} && \multicolumn{3}{c}{$\bar{u}\bar{s}cb$}	\\
\cline{3-5}\cline{7-9}
	&&	$V^{C}$	&	$V^{CS}$  & 	$l$	&&	$V^{C}$	&	$V^{CS}$  & 	$l$	\\
\hline
$(1,2)$
    &&	76.45	&	-104.4	  &	 0.820   && 	-10.65		&	-22.51	& 0.855  \\
$(1,3)$
    && 	119.0	&	-49.34	  &	 0.656   && 	204.4		&	-92.06	& 0.585	\\
$(1,4)$
    && 	117.4	&	-18.63    &  0.638	 && 	61.61		&	-11.98	& 0.736	\\
$(2,3)$
    && 	119.0	&	-49.34	  &	 0.656   && 	22.51		&	-29.10	& 0.673	\\
$(2,4)$
    && 	117.4	&	-18.63	  &	 0.638   && 	22.44		&	-39.97	& 0.440	\\
$(3,4)$
    && 	-2.817	&	-7.717	  &  0.530   && 	-14.28		&	-2.195	& 0.655	\\
\hline
Total
    && 	546.4	&	-248.0	  & 	&& 	286.0		&	-197.8	&	\\
\hline
\hline
\label{TetraSeparation1}
\end{tabular}
\end{table}

\begin{table}[h]

\caption{Color interaction factor, $-\langle \lambda^{c}_{i} \lambda^{c}_{j} \rangle$), and CS interaction factor, $-\langle \lambda^{c}_{i} \lambda^{c}_{j} \boldsymbol{\sigma}_i \cdot \boldsymbol{\sigma}_j \rangle$, for the $\bar{u}\bar{d}cb$ with $(I,S)=(0,0)$ and the $\bar{u}\bar{s}cb$ with $(I,S)=(1/2,0)$ states.}	
\centering
\begin{tabular}{ccccccc}
\hline
\hline
\multirow{2}{*}{Quark pair}	&& \multicolumn{2}{c}{$-\langle \lambda^{c}_{i} \lambda^{c}_{j} \rangle$} && \multicolumn{2}{c}{$-\langle \lambda^{c}_{i} \lambda^{c}_{j} \boldsymbol{\sigma}_i \cdot \boldsymbol{\sigma}_j \rangle$}	\\
\cline{3-4}\cline{6-7}
   &&$\bar{u}\bar{d}cb$&$\bar{u}\bar{s}cb$&&$\bar{u}\bar{d}cb$&$\bar{u}\bar{s}cb$	\\
\hline
$(1,2)$ &&	0.9150  &  -0.074    &&	-4.206  &  -1.493        \\
$(1,3)$ &&	2.209  &  4.114     &&	-7.105  &  -12.59        \\
$(1,4)$ &&	2.209  &  1.294     &&	-7.105  &  -4.347     \\
$(2,3)$ &&	2.209  &  1.294     &&	-7.105  &  -4.347     \\
$(2,4)$ &&	2.209  &  4.114     &&	-7.105  &  -12.59     \\
$(3,4)$ &&	0.9150  &  -0.074    &&	-4.206  &  -1.493     \\
\hline 
\hline
\label{C-CS-Factors-udcb-uscb}
\end{tabular}
\end{table}

\begin{widetext}

\begin{table}[h]

\caption{The matrices of $\lambda^{c}_{i} \lambda^{c}_{j}$ and $\lambda^{c}_{i} \lambda^{c}_{j} \boldsymbol{\sigma}_i \cdot \boldsymbol{\sigma}_j$ for each quark pair $(i,j)$ in the total spin 0 tetraquark configuration. The color bases of $\lambda^{c}_{i} \lambda^{c}_{j}$ are defined as $|1\rangle_C \equiv | \{\bar{q}_1 \bar{q}_2\}^\mathbf{3} \{Q_3 Q_4\}^\mathbf{\bar{3}} \rangle$ and $|2\rangle_C \equiv | \{\bar{q}_1 \bar{q}_2 \}^\mathbf{\bar{6}} \{Q_3 Q_4\}^\mathbf{6} \rangle$. The CS bases of $\lambda^{c}_{i} \lambda^{c}_{j} \boldsymbol{\sigma}_i \cdot \boldsymbol{\sigma}_j$ are defined as $|1\rangle_{CS} \equiv | \{\bar{q}_1\bar{q}_2\}^{\mathbf{3}}_1 \{Q_3 Q_4\}^{\mathbf{\bar{3}}}_1 \rangle$, $|2\rangle_{CS} \equiv | \{\bar{q}_1 \bar{q}_2\}^{\mathbf{3}}_0 \{Q_3 Q_4\}^{\mathbf{\bar{3}}}_0 \rangle$, $|3\rangle_{CS} \equiv | \{\bar{q}_1 \bar{q}_2\}^{\mathbf{\bar{6}}}_1 \{Q_3 Q_4\}^{\mathbf{6}}_1 \rangle$, and $|4\rangle_{CS} \equiv | \{\bar{q}_1 \bar{q}_2 \}^{\mathbf{\bar{6}}}_0 \{Q_3 Q_4\}^{\mathbf{6}}_0 \rangle$.}
\centering
\begin{tabular}{ccccccc}
\hline
\hline
Quark pair	&& $(1,2)$ \& $(3,4)$ & $(1,3)$ \& $(2,4)$  & $(1,4)$ \& $(2,3)$	\\
\hline
        $-\lambda^{c}_{i} \lambda^{c}_{j}$
        &&
        $\left(\begin{array}{cc}
          \frac{8}{3}   & 0 \\
          0             & -\frac{4}{3}
        \end{array}\right)$
        &
       $\left(\begin{array}{cc}
          \frac{4}{3}  & 2\sqrt{2} \\
          2\sqrt{2}     & \frac{10}{3}
        \end{array}\right)$ 
        &
        $\left(\begin{array}{cc}
          \frac{4}{3}  & -2\sqrt{2} \\
          -2\sqrt{2}     & \frac{10}{3}
        \end{array}\right)$         \\
        $-\lambda^{c}_{i} \lambda^{c}_{j} \boldsymbol{\sigma}_i \cdot \boldsymbol{\sigma}_j$
        &&
        $\left(\begin{array}{cccc}
          \frac{8}{3}  & 0  & 0 & 0 \\
          0  & -8    & 0 & 0 \\
          0 & 0   & -\frac{4}{3}   & 0 \\
          0 & 0 & 0 & 4
        \end{array}\right)$
        &
        $\left(\begin{array}{cccc}
          -\frac{8}{3}  & -\frac{4}{\sqrt{3}}  & -4 \sqrt{2} & -2 \sqrt{6} \\
          -\frac{4}{\sqrt{3}}  & 0    & -2 \sqrt{6} & 0 \\
          -4 \sqrt{2} & -2 \sqrt{6}   & -\frac{20}{3}   & -\frac{10}{\sqrt{3}} \\
          -2 \sqrt{6} & 0 & -\frac{10}{\sqrt{3}} & 0
        \end{array}\right)$
        &
        $\left(\begin{array}{cccc}
          -\frac{8}{3}  & \frac{4}{\sqrt{3}}  & 4 \sqrt{2} & -2 \sqrt{6} \\
          \frac{4}{\sqrt{3}}  & 0    & -2 \sqrt{6} & 0 \\
          4 \sqrt{2} & -2 \sqrt{6}   & -\frac{20}{3}   & \frac{10}{\sqrt{3}} \\
          -2 \sqrt{6} & 0 & \frac{10}{\sqrt{3}} & 0
        \end{array}\right)$ \\
\hline
\hline
\label{Lambda_CS_Matrix}
\end{tabular}
\end{table}

\end{widetext}

First, it is crucial to emphasize that the primary contribution of the color states to the ground state differs significantly between the $\bar{u}\bar{d}cb$ and $\bar{u}\bar{s}cb$ configurations.
Specifically, for the $\bar{u}\bar{d}cb$, the dominant state is the color triplet state, $\left| \left\{ \bar{q}_1 \bar{q}_2 \right\}^{\mathbf{3}} \left\{ Q_3 Q_4 \right\}^{\mathbf{\bar{3}}} \right\rangle$, whereas for the $\bar{u}\bar{s}cb$, it is the color sextet state, $\left| \left\{ \bar{q}_1 \bar{q}_2 \right\}^{\mathbf{\bar{6}}} \left\{ Q_3 Q_4 \right\}^{\mathbf{6}} \right\rangle$.
This distinction is directly inferred from the respective color interaction factors, $-\left\langle \lambda^{c}_{1} \lambda^{c}_{2} \right\rangle$, as detailed in Table~\ref{C-CS-Factors-udcb-uscb}, along with the matrix elements provided in Table~\ref{Lambda_CS_Matrix}:
\begin{widetext}

\begin{eqnarray}
- \left\langle \Psi^{\bar{u}\bar{d}cb}_{1} \right| \lambda^{c}_{1} \lambda^{c}_{2} \left| \Psi^{\bar{u}\bar{d}cb}_{1} \right\rangle
&&=
\sum_{k}^{k_{\rm max}} \sum_{k'}^{k'_{\rm max}} \bigg[ \bigg\{ C^{\bar{u}\bar{d}cb}_1 (1,k) C^{\bar{u}\bar{d}cb}_1 (1,k') + C^{\bar{u}\bar{d}cb}_2 (1,k) C^{\bar{u}\bar{d}cb}_2 (1,k') \bigg\} \left( \frac{8}{3} \right)
\nonumber \\
&&\hspace{2.2cm}+ \bigg\{ C^{\bar{u}\bar{d}cb}_3 (1,k) C^{\bar{u}\bar{d}cb}_3 (1,k') + C^{\bar{u}\bar{d}cb}_4 (1,k) C^{\bar{u}\bar{d}cb}_4 (1,k') \bigg\} \left( -\frac{4}{3} \right) \bigg] \left\langle \phi_k | \phi_{k'} \right\rangle
\nonumber \\
&&\hspace{-0.0cm}=
\sum_k^{k_{\rm max}} \bigg[ \left\{ \left( C^{\bar{u}\bar{d}cb}_1 (1,k) \right)^2 + \left( C^{\bar{u}\bar{d}cb}_2 (1,k) \right)^2 \right\} \left( \frac{8}{3} \right)
\nonumber \\
&&\hspace{2.2cm}+ \left\{ \left( C^{\bar{u}\bar{d}cb}_3 (1,k) \right)^2 + \left( C^{\bar{u}\bar{d}cb}_4 (1,k) \right)^2 \right\} \left( -\frac{4}{3} \right) \bigg]
\nonumber \\
&&\hspace{-0.0cm}=
\sum_k^{k_{\rm max}} \bigg[ \left\{ \left( C^{\bar{u}\bar{d}cb}_1 (1,k) \right)^2 + \left( C^{\bar{u}\bar{d}cb}_2 (1,k) \right)^2 + \left( C^{\bar{u}\bar{d}cb}_3 (1,k) \right)^2 + \left( C^{\bar{u}\bar{d}cb}_4 (1,k) \right)^2 \right\} \left( - \frac{4}{3} \right)
\nonumber \\
&&\hspace{2.2cm}+ \left\{ \left( C^{\bar{u}\bar{d}cb}_1 (1,k) \right)^2 + \left( C^{\bar{u}\bar{d}cb}_2 (1,k) \right)^2 \right\} \left( \frac{12}{3} \right) \bigg]
\nonumber \\
&&\hspace{-0.0cm}=
-\frac{4}{3} + \sum_k^{k_{\rm max}} \left\{ \left( C^{\bar{u}\bar{d}cb}_1 (1,k) \right)^2 + \left( C^{\bar{u}\bar{d}cb}_2 (1,k) \right)^2 \right\} \left( \frac{12}{3} \right) = 0.9150\,,
\label{udcbColorExpand}
\end{eqnarray}

\end{widetext}
where $\left| \Psi^{\bar{u}\bar{d}cb}_{1} \right\rangle$ is the ground state of the $\bar{u}\bar{d}cb$, which is given from Eq.~(\ref{H_eigenstate}).
In Eq.~(\ref{udcbColorExpand}), the factor $\frac{8}{3}$ comes from the color triplet state, the $-\frac{4}{3}$ from the color sextet state, and there is no off-diagonal element between these two color states, which can be inferred from the color matrix for the $(1,2)$ pair in Table~\ref{Lambda_CS_Matrix}.
The value of $- \left\langle \lambda^{c}_{1} \lambda^{c}_{2} \right\rangle
$, $0.9150$, in Eq.~(\ref{udcbColorExpand}), is given in Table~\ref{C-CS-Factors-udcb-uscb}.

From Eq.~(\ref{udcbColorExpand}), we immediately obtain the following relation:
\begin{eqnarray}
\sum_k^{k_{\rm max}} \left\{ \left( C^{\bar{u}\bar{d}cb}_1 (1,k) \right)^2 + \left( C^{\bar{u}\bar{d}cb}_2 (1,k) \right)^2 \right\} = 0.562083\,.
\nonumber \\
\label{Color_Ample_udcb}
\end{eqnarray}
Correspondingly, we also obtain another relation for the color sextet contribution as follows:
\begin{eqnarray}
\sum_k^{k_{\rm max}} \left\{ \left( C^{\bar{u}\bar{d}cb}_3 (1,k) \right)^2 + \left( C^{\bar{u}\bar{d}cb}_4 (1,k) \right)^2 \right\} = 0.437917\,.
\nonumber \\
\end{eqnarray}
Likewise, one can obtain the sum of probability amplitudes for the $\bar{u}\bar{s}cb$ as follows.
\begin{eqnarray}
\sum_k^{k_{\rm max}} \left( C^{\bar{u}\bar{s}cb}_1 (1,k) \right)^2 + \left( C^{\bar{u}\bar{s}cb}_2 (1,k) \right)^2 = 0.314833\,,
\nonumber \\
\sum_k^{k_{\rm max}} \left( C^{\bar{u}\bar{s}cb}_3 (1,k) \right)^2 + \left( C^{\bar{u}\bar{s}cb}_4 (1,k) \right)^2 = 0.685167\,.
\label{Color_Ample_uscb}
\end{eqnarray}
From Eqs.~(\ref{Color_Ample_udcb}) and (\ref{Color_Ample_uscb}), it is obvious that the dominant color state for the $\bar{u}\bar{d}cb$ is triplet, whereas for the $\bar{u}\bar{s}cb$, it shifts to the sextet state.
This leads to a considerable difference in the binding energy between these two tetraquark states, which results in the relative attraction of the confinement part in the $\bar{u}\bar{s}cb$, as shown from Table~\ref{TetraSeparation1}.
Replacing the $d$ quark with the $s$ quark in the $\bar{u}\bar{d}cb$ configuration breaks the symmetry in the light quark sector, leading to a shift in the dominant color state from the triplet to the sextet.
The triplet state is dominant for all other bound tetraquark states, except for the $\bar{u}\bar{s}cb$ configuration, listed in Table~\ref{result}.

Another notable feature of tetraquark states is that the symmetry imposed by the Pauli principle is required to construct the total wave function and is reflected in the values of the $-\langle \lambda^{c}_{i} \lambda^{c}_{j} \rangle$ and $-\langle \lambda^{c}_{i} \lambda^{c}_{j} \boldsymbol{\sigma}_i \cdot \boldsymbol{\sigma}_j \rangle$, as presented in Table~\ref{C-CS-Factors-udcb-uscb}.
From Table~\ref{Lambda_CS_Matrix}, it can be found that the matrices, $-\lambda^{c}_{i} \lambda^{c}_{j}$ and $-\lambda^{c}_{i} \lambda^{c}_{j} \boldsymbol{\sigma}_i \cdot \boldsymbol{\sigma}_j$, are identical for the pairs $(1,2)$ and $(3,4)$, $(1,3)$ and $(2,4)$, $(1,4)$ and $(2,3)$, respectively.
Thus, for tetraquark states, $-\langle \lambda^{c}_{i} \lambda^{c}_{j} \rangle$ and $-\langle \lambda^{c}_{i} \lambda^{c}_{j} \boldsymbol{\sigma}_i \cdot \boldsymbol{\sigma}_j \rangle$ should also follow these symmetry rules.
However, unlike the $\bar{u}\bar{s}cb$, the total wave function of the $\bar{u}\bar{d}cb$ should be antisymmetric under the permutation $(12)$ due to the Pauli principle.
This requirement imposes an additional permutation symmetry on $-\langle \lambda^{c}_{i} \lambda^{c}_{j} \rangle$ and $-\langle \lambda^{c}_{i} \lambda^{c}_{j} \boldsymbol{\sigma}_i \cdot \boldsymbol{\sigma}_j \rangle$ for the $\bar{u}\bar{d}cb$ as follows:
\begin{eqnarray}
&&\left\langle \Psi^{\bar{u}\bar{d}cb}_{1} \right| \lambda^{c}_{1} \lambda^{c}_{3} \left| \Psi^{\bar{u}\bar{d}cb}_{1} \right\rangle
\nonumber \\
&&=
\left\langle \Psi^{\bar{u}\bar{d}cb}_{1} \right| (12)^{-1} (12) \lambda^{c}_{1} \lambda^{c}_{3} (12)^{-1} (12) \left| \Psi^{\bar{u}\bar{d}cb}_{1} \right\rangle
\nonumber \\
&&=
\left\langle \Psi^{\bar{u}\bar{d}cb}_{1} \right| \lambda^{c}_{2} \lambda^{c}_{3} \left| \Psi^{\bar{u}\bar{d}cb}_{1} \right\rangle,
\nonumber \\
&&\left\langle \Psi^{\bar{u}\bar{d}cb}_{1} \right| \lambda^{c}_{1} \lambda^{c}_{4} \left| \Psi^{\bar{u}\bar{d}cb}_{1} \right\rangle
\nonumber \\
&&=
\left\langle \Psi^{\bar{u}\bar{d}cb}_{1} \right| (12)^{-1} (12) \lambda^{c}_{1} \lambda^{c}_{4} (12)^{-1} (12) \left| \Psi^{\bar{u}\bar{d}cb}_{1} \right\rangle
\nonumber \\
&&=
\left\langle \Psi^{\bar{u}\bar{d}cb}_{1} \right| \lambda^{c}_{2} \lambda^{c}_{4} \left| \Psi^{\bar{u}\bar{d}cb}_{1} \right\rangle,
\label{udcb_symmetry}
\end{eqnarray}
where $\Psi^{\bar{u}\bar{d}cb}_{1}$ is defined as $\Psi_{1}$ in Eq.~(\ref{H_eigenstate}) but satisfies the symmetry constraints for the $\bar{u}\bar{d}cb$.
Therefore, as shown in Table~\ref{C-CS-Factors-udcb-uscb}, this symmetry explains why the values only for the $(1,3)$ and $(2,3)$ pairs are identical only for the $\bar{u}\bar{d}cb$ configuration.

\begin{table}[h]

\caption{``Non-mixing'' and ``Mixing'' components of $V^C$ part, presented in Table~\ref{TetraSeparation1}, for the $\bar{u}\bar{d}cb$ with $(I,S) = (0,0)$ and the $\bar{u}\bar{s}cb$ with $(I,S) = (1/2,0)$. All values are given in MeV unit.}
\centering
\begin{tabular}{cccccc}
\hline
\hline
\multirow{2}{*}{Quark Pair} &&&	\multicolumn{3}{c}{$\bar{u}\bar{d}cb$}      \\
\cline{4-6}
&&&Non-mixing&Mixing&Subtotal\\
\hline
    	$(1,2)$			&&&	104.2    &  -27.75   &   76.45       \\
    	$(1,3)$			&&&	171.1    &	-52.13   &   119.0       \\
    	$(1,4)$			&&&	162.6    &	-45.25   &   117.4       \\
    	$(2,3)$			&&&	171.1    &	-52.13   &   119.0       \\
    	$(2,4)$			&&&	162.6    &	-45.25   &   117.4       \\
    $(3,4)$			&&&	41.79    &	-44.60   &   -2.817       \\
\hline
        Total		&&&	813.4    &	-267.1   &   546.4       \\
\hline  
\hline
\multirow{2}{*}{Quark Pair} &&&	\multicolumn{3}{c}{$\bar{u}\bar{s}cb$}      \\
\cline{4-6}
&&&Non-mixing&Mixing&Subtotal\\
\hline
    	$(1,2)$			&&&	-9.042   &  -1.610   &   -10.65       \\
    	$(1,3)$			&&&	248.7    &	-44.30   &   204.4       \\
    	$(1,4)$			&&&	123.8    &	-62.16   &   61.61       \\
    	$(2,3)$			&&&	103.9    &	-81.40   &   22.51       \\
    	$(2,4)$			&&&	80.51    &	-58.07   &   22.44       \\
    $(3,4)$			&&&	-5.822   &	-8.460   &   -14.28       \\
\hline
        Total		&&&	542.0    &	-256.0   &   286.0       \\
\hline  
\hline
\label{Detailed_VC_udcb_and_uscb}
\end{tabular}
\end{table}

For detailed numerical analysis of the confinement potential $V^{C}$ part, we refer to Table~\ref{Detailed_VC_udcb_and_uscb}.
This analysis utilizes the quark pair sizes listed in Table~\ref{TetraSeparation1} and the color interaction factor, $-\langle \lambda^{c}_{i} \lambda^{c}_{j} \rangle$, shown in Table~\ref{C-CS-Factors-udcb-uscb}.

For the $(1,2)$ and $(3,4)$ pairs, the magnitudes of the non-mixing components for the $\bar{u}\bar{s}cb$ are much smaller and the signs are reversed compared to those for the $\bar{u}\bar{d}cb$.
This difference arises because the magnitude of $-\langle \lambda^{c}_{i} \lambda^{c}_{j} \rangle$ is smaller by an order of magnitude and the sign is reversed for the $\bar{u}\bar{s}cb$ compared to the $\bar{u}\bar{d}cb$.
Even though the sizes of these quark pairs for the $\bar{u}\bar{s}cb$ are larger, it does not significantly affect the magnitude of the non-mixing components, as in the comparison of the $(1,2)$ pair for the $V^C$ part between the $\bar{u}\bar{s}cb$ and $\bar{u}\bar{d}cb$ states discussed in Sec.~\ref{SecIIIA}.

For the $(1,3)$ pair, the non-mixing component of the $\bar{u}\bar{s}cb$ is larger than that of the $\bar{u}\bar{d}cb$.
This is primarily due to the larger value of $-\langle \lambda^{c}_{i} \lambda^{c}_{j} \rangle$ for the $\bar{u}\bar{s}cb$.
Although the spatial part is contrary to this, it does not exceed the increase in the value of $-\langle \lambda^{c}_{i} \lambda^{c}_{j} \rangle$ because the spatial part for the $\bar{u}\bar{s}cb$ is evaluated at a slightly smaller size compared to the $\bar{u}\bar{d}cb$.
Conversely, for the $(2,4)$ pair, the non-mixing component of the $\bar{u}\bar{s}cb$ is smaller than that of the $\bar{u}\bar{d}cb$ even though $-\langle \lambda^{c}_{i} \lambda^{c}_{j} \rangle$ is the same as for the $(1,3)$ pair.
This is due to the larger size difference of the $(2,4)$ pair between the two tetraquark states compared to the $(1,3)$ pair.
Additionally, the decrease in value of the spatial part for the $(2,4)$ pair is very appreciable compared to the $(1,3)$ pair because the size of the $(2,4)$ pair for the $\bar{u}\bar{s}cb$ is much closer to 0.32 fm, where the spatial part of $V^C$ nearly goes to zero.

For both the $(1,4)$ and $(2,3)$ pairs, despite the larger sizes for the $\bar{u}\bar{s}cb$, the magnitudes of the non-mixing components are smaller.
This is primarily because $-\langle \lambda^{c}_{i} \lambda^{c}_{j} \rangle$ for the $\bar{u}\bar{s}cb$ is nearly half of that for the $\bar{u}\bar{d}cb$.
In the case of the $(2,3)$ pair, this result is evident because the increase in size for the $\bar{u}\bar{s}cb$ is slight. For the $(1,4)$ pair, the increase in size for the $\bar{u}\bar{s}cb$ only reduces the difference in the non-mixing components between the two tetraquark states.

\begin{table}[h]

\caption{Similar to Table~\ref{Detailed_VC_udcb_and_uscb} but for $V^{CS}$ part in Table~\ref{TetraSeparation1}.}	
\centering
\begin{tabular}{cccccc}
\hline
\hline
\multirow{2}{*}{Quark Pair} &&&	\multicolumn{3}{c}{$\bar{u}\bar{d}cb$}      \\
\cline{4-6}
&&&Non-mixing&Mixing&Subtotal\\
\hline
    	$(1,2)$			&&&	-83.63    &  -20.77 &   -104.4       \\
    	$(1,3)$			&&&	-39.58    &	-9.755  &   -49.34       \\
    	$(1,4)$			&&&	-15.25    &	-3.377  &   -18.63       \\
    	$(2,3)$			&&&	-39.58    &	-9.755  &   -49.34       \\
    	$(2,4)$			&&&	-15.25    &	-3.377  &   -18.63       \\
    $(3,4)$			&&&	-5.476    &	-2.241  &   -7.717       \\
\hline
        Total		&&&	-198.8    &	-49.28   &   -248.0       \\
\hline  
\hline
\multirow{2}{*}{Quark Pair} &&&	\multicolumn{3}{c}{$\bar{u}\bar{s}cb$}      \\
\cline{4-6}
&&&Non-mixing&Mixing&Subtotal\\
\hline
    	$(1,2)$			&&&	-14.54   &  -7.962    &   -22.51       \\
    	$(1,3)$			&&&	-83.81   &	-8.242    &   -92.06       \\
    	$(1,4)$			&&&	-7.494   &	-4.489    &   -11.98       \\
    	$(2,3)$			&&&	-16.03   &	-13.07    &   -29.10       \\
    	$(2,4)$			&&&	-35.39   &	-4.581    &   -39.97       \\
    $(3,4)$			&&&	-1.153   &	-1.042   &   -2.195       \\
\hline
        Total		&&&	-158.4    &	-39.39   &   -197.8       \\
\hline  
\hline
\label{Detailed_VCS_udcb_and_uscb}
\end{tabular}
\end{table}

We now turn our attention to the hyperfine potential $V^{CS}$ part shown in Table~\ref{TetraSeparation1}. For this analysis, we refer to Table~\ref{Detailed_VCS_udcb_and_uscb}. This analysis utilizes the quark pair sizes listed in Table~\ref{TetraSeparation1} and the CS interaction factor, $-\langle \lambda^{c}_{i} \lambda^{c}_{j} \boldsymbol{\sigma}_i \cdot \boldsymbol{\sigma}_j \rangle$, presented in Table~\ref{C-CS-Factors-udcb-uscb}.

For the $(1,2)$ and $(3,4)$ pairs, the magnitudes of the non-mixing components for the $\bar{u}\bar{s}cb$ are smaller than those for the $\bar{u}\bar{d}cb$.
This reduction is due to the fact that the magnitude of $-\langle \lambda^{c}_{i} \lambda^{c}_{j} \boldsymbol{\sigma}_i \cdot \boldsymbol{\sigma}_j \rangle$ for the $\bar{u}\bar{s}cb$ is smaller by almost a factor of 1/3.
Additionally, the sizes of these quark pairs for the $\bar{u}\bar{s}cb$ are larger.
Furthermore, the hyperfine potential is inversely proportional to the quark pair mass involved. Therefore, the presence of a strange quark in the $\bar{u}\bar{s}cb$ configuration leads to a more significant decrease in the magnitude of the non-mixing component for the $(1,2)$ pair compared to the $(3,4)$ pair.
This can be obtained from a rough estimate provided in Section~\ref{SecIIIA}. From Table~\ref{Detailed_VCS_udcb_and_uscb}, the ratio of the magnitude of the non-mixing term of the $\bar{u}\bar{s}cb$ to that of the $\bar{u}\bar{d}cb$ is 0.17 for the $(1,2)$ pair.
By first obtaining the ratio of the inverse of sizes and the ratio of the values of $-\langle \lambda^{c}_{i} \lambda^{c}_{j} \boldsymbol{\sigma}_i \cdot \boldsymbol{\sigma}_j \rangle$, which evaluates as 0.34, we find this value is roughly compensated by the ratio of the mass of a light quark to that of a strange quark, which is 0.52, using the quark masses from Eq.~(\ref{FitParameters}).

For the $(1,3)$ and $(2,4)$ pairs, the magnitudes of the non-mixing components for the $\bar{u}\bar{s}cb$ are larger than those for the $\bar{u}\bar{d}cb$.
This is obvious not only because the sizes of these quark pairs for the $\bar{u}\bar{s}cb$ are smaller but also because the magnitude of $-\langle \lambda^{c}_{i} \lambda^{c}_{j} \boldsymbol{\sigma}_i \cdot \boldsymbol{\sigma}_j \rangle$ for the $\bar{u}\bar{s}cb$ are larger.

For the $(1,4)$ and $(2,3)$ pairs, the magnitudes of the non-mixing components for the $\bar{u}\bar{s}cb$ become smaller, which can be accounted by the reasons contrary to those in the analysis of the $(1,3)$ and $(2,4)$ pairs.

From our quantitative analysis for the confinement and hyperfine potential parts, we find that the substantial difference in the binding energy between these two tetraquark states primarily results from the relative attraction of the confinement part rather than the hyperfine potential part in the $\bar{u}\bar{s}cb$.
This is fundamentally traceable to the shift of the dominant color state from the triplet to the sextet state in the $\bar{u}\bar{s}cb$ state.

It should also be noted that the dominant color contribution for the $\bar{u}\bar{s}bb$ comes from the triplet state, which is different from that for the $\bar{u}\bar{s}cb$.
This difference arises from the strong interactions within the $bb$ pair, similar to the case of the $\bar{u}\bar{d}bb$.
For instance, in the $\bar{u}\bar{s}bb$ with spin 1 state, the sum of the probability amplitudes of the triplet states is 0.913, which is close to the corresponding value of 0.968 for the $\bar{u}\bar{d}bb$ with spin 1 state.
Furthermore, the size of the $bb$ pair is quite similar in both cases, being 0.293 fm in the $\bar{u}\bar{d}bb$ and 0.319 fm in the $\bar{u}\bar{s}bb$.
Consequently, the strong interaction within the $bb$ pair provides the dominant contribution from the triplet state, overwhelming the contribution from the sextet state.
On the other hand, for the $\bar{u}\bar{d}cc$ with spin 1 state, where the size of the heavy quark pair($cc$) is 0.64 fm as presented in Ref.~\cite{Noh:2023zoq}, the sum of the probability amplitudes from the triplet state is 0.662.

\section*{Acknowledgements}
This work was supported by the Korea National Research Foundation under the grant number 2021R1A2C1009486(NRF).

\end{document}